\documentclass[prd,aps,a4paper,twocolumn,nofootinbib]{revtex4-2}

\usepackage[utf8]{inputenc}
\usepackage[normalem]{ulem}

\usepackage{graphicx,psfrag}
\usepackage{mathrsfs}
\usepackage{amsmath,amsfonts,amssymb}
\usepackage{multirow}
\usepackage{comment}
\usepackage{bm}
\usepackage{epstopdf}
\usepackage{color}
\usepackage{mathrsfs}
\usepackage{comment}
\usepackage{multirow}
\usepackage{hyperref}
\usepackage{diagbox} 
\usepackage{amssymb,amsthm,wasysym}
\usepackage{capt-of}
\usepackage{physics}

\def\mbar{\overline{m}}
\def\hr{{\hat{r}}}
\def\ha{{\hat{a}}}
\def\rmd{{\rm d}}

\DeclareMathOperator{\sgn}{sgn}

\newcommand{\be}{\begin{equation}}
\newcommand{\ee}{\end{equation}}

\newcommand{\RN}[1]{%
  \textup{\uppercase\expandafter{\romannumeral#1}}%
}

\definecolor{gray}{rgb}{0.5,0.5,0.5}
\definecolor{cyan}{rgb}{0,0.9,0.9}
\definecolor{orange}{rgb}{0.9,0.5,0}
\definecolor{magenta}{rgb}{1,0,1}
\definecolor{purple}{rgb}{0.8,0.4,0.8}
\definecolor{darkgreen}{rgb}{0,.8,0}
\definecolor{turquoise}{rgb}{0.25,0.88,0.82}

\begin{document}

\interfootnotelinepenalty=10000
\raggedbottom

\title{Adiabatic equatorial inspirals of a spinning body into a Kerr black hole}

\author{Viktor Skoup\'y$^{1,\,2}$}
\author{Georgios Lukes-Gerakopoulos$^1$}
\affiliation{${}^1$ Astronomical Institute of the Czech Academy of Sciences, Bo\v{c}n\'{i} II 1401/1a, CZ-141 00 Prague, Czech Republic}   
\affiliation{${}^2$Institute of Theoretical Physics, Faculty of Mathematics and Physics, Charles University, CZ-180 00 Prague, Czech Republic}

\begin{abstract}
The detection of gravitational waves from Extreme mass Ratio Inspirals (EMRIs) by the future space-based gravitational-wave detectors demands the generation of accurate enough waveform templates. Since the spin of the smaller secondary body cannot be neglected for the detection and parameter estimation of EMRIs, we study its influence on the phase of the gravitational waves from EMRIs with spinning secondary. We focus on generic eccentric equatorial orbits around a Kerr black hole. To model the spinning secondary object, we use the Mathisson-Papapetrou-Dixon equations in the pole-dipole approximation. Furthermore, we linearize in spin the orbital variables and the gravitational-wave fluxes from the respective orbits. We obtain these fluxes by using the Teukolsky formalism in the frequency domain. We derive the evolution equations for the spin induced corrections to the adiabatic evolution of an inspiral. Finally, through their numerical integration we find the gravitational-wave phase shift between an inspiral of a spinning and a non-spinnig body. 
\end{abstract}

\maketitle

\section{Introduction}
\label{sec:intro}

Extreme mass ratio inspirals (EMRIs) are promising sources for the future space-based of gravitational wave (GW) detectors such as the Laser Interferometer Space Antenna (LISA) \citep{LISA,Babak:2017}. These systems consist of a primary supermassive black hole and a secondary much lighter compact object such as a neutron star or a black hole. In an EMRI the mass ratio $q=\mu/M$ of the secondary mass $\mu$ and the primary mass $M$ is expected to lie between $10^{-7}$ and $10^{-4}$. Due to the gravitational radiation reaction the secondary object is slowly inspiraling into the primary while it radiates gravitational waves. The detection of EMRIs will provide the opportunity to study strong gravitational fields around supermassive black holes lying at the centre of galaxies and to test general relativity. 

The mHz GW bandwidth that EMRIs are emitting is expected to be rich in GW sources. To overcome the fact that signals from various sources will overlap during their detection by LISA, matched filtering is planned to be employed, i.e. the detected signal will be compared with large number of GW templates covering the estimated parameter space \cite{LISA}. The use of templates will not only allow the detection of EMRI signals, but it also be employed for the parameter estimation of these systems. To get these estimations adequately enough, we need to generate waveform templates whose phases are accurate up to fractions of radians.

To achieve such accuracy, a series of techniques can be employed. The backbone of them is that the system is treated as the motion of a secondary object in the background spacetime of the primary object. Hence, to model the GW phase, we need first to find the trajectory of the secondary $z^\mu$. The secondary is perturbing the background spacetime and the gravitational self force drives the secondary away from the trajectory which it would follow without this perturbation\footnote{This unperturbed trajectory is a geodesic orbit for a non-spinning secondary, while for a spinning secondary, the trajectory can be provided by the Mathisson-Papapetrou-Dixon equations.}. To find this self-force, perturbation theory is used. Namely, the exact metric is expanded in the terms of the mass ratio as
\begin{equation}
    g^{\rm exact}_{\mu\nu} = g_{\mu\nu} + h^{(1)}_{\mu\nu} + h^{(2)}_{\mu\nu} + \order{q^3} \; ,
\end{equation}
where $g_{\mu\nu}$ is the background metric, which in our case is the Kerr one, $h^{(1)}_{\mu\nu} = \order{q}$ is the first order perturbation and $h^{(2)}_{\mu\nu} = \order{q}^2$ is the second order perturbation. $h^{(n)}_{\mu\nu}$ are found by expanding the Einstein's equations in the mass ratio with the source constructed from the secondary body and solving order-by-order \citep{Barack:2019,Pound:2021}. The parts of the metric perturbation are then used to construct the first and second-order self force
\begin{equation}
    \frac{{\rm D}^2 z^\mu}{\dd \tau^2} = q f^\mu_{(1)} + q^2 f^\mu_{(2)} + \order{\sigma^3}\, ,
\end{equation}
where $\tau$ is the proper time, $f^\mu_{(1)}$ is constructed from $h^{(1)}_{\mu\nu}$ and the secondary's spin-curvature coupling, while $f^\mu_{(2)}$ is constructed from $h^{(2)}_{\mu\nu}$ \citep{Gralla:2008,Pound:2010,Pound:2021}.

Because the radiation-reaction is of the order $\order{q}$, its effects act on a much larger timescale than is the orbital timescale. Actually, the secondary makes $\order{q^{-1}}$ orbits due to the radiation reaction, before it plunges into the primary.  Thanks to this timescale difference we can use the so-called \emph{two-timescale approximation} \citep{Hinderer:2008}. In this approximation, the coordinates are transformed to angle-like variables $q_\mu$, which can be expanded in the mass ratio as\footnote{In fact, the expansion contains also term proportional to $q^{-1/2}$ caused by the orbital resonances, but here we neglect it for simplicity.} 
\begin{equation}
    q_\mu(t) = \frac{1}{q} q^{(0)}_\mu(q\,t) + q^{(1)}_\mu(q\,t) + \order{q}\, ,
\end{equation}
where $t$ is the evolution parameter. The first term $q^{(0)}_\mu(q\,t)$ is called \emph{adiabatic} term and can be calculated only from the time-averaged dissipative part of the first-order self force. The second term $q^{(1)}_\mu(q\,t)$, which is called the first-order \emph{post-adiabatic} term, is constructed from the oscillating dissipative and conservative parts of the first-order self force, the averaged dissipative part of the second-order self force and the contribution from the spin of the secondary body. These angle variables are directly related to the phases of the GW. The adiabatic term for generic orbits around a Kerr black hole was calculated only recently \citep{Fujita:2020,Hughes:2021,Chua:2021,Katz:2021} and, so far, the post-adiabatic term with the first-order self-force was calculated for spinning secondary only for quasicircular orbits in the Schwarzschild spacetime \cite{Mathews2021} and for non-spinning secondary for equatorial orbits in the Schwarzschild \citep{Osburn:2016} and Kerr \citep{Lynch:2021} spacetime, while the full first and second-order self-force for quasicircular orbits in the Schwarzschild spacetime was calculated in \citep{Wardell:2021}.

The error in the adiabatic term must be less than the mass ratio to obtain the sub-radian precision. It has been proven for a non-spinning secondary \citep{Mino:2003,Isoyama:2019}, but also for a spinning secondary \citep{Akcay:2020} that the time-averaged dissipative part of the self-force can be reconstructed from the time-averaged energy and angular momentum fluxes calculated at infinity and at the horizon of the primary black hole. Therefore, for the calculations in the adiabatic order, we do not need to calculate the perturbation $h^{(1)}_{\mu\nu}$ in the vicinity of the secondary body, but we only need to find the aforementioned GW fluxes. These fluxes were calculated for generic orbits of non-spining bodies around a Kerr black hole in \citep{Drasco:2005kz}, for circular orbits of spinning bodies around a Schwarzschild and a Kerr black hole in \citep{Harms:2016a,Harms:2016b,Lukes-Gerakopoulos:2017,Nagar:2019,Akcay:2020,Piovano:2020} and, finally, for eccentric equatorial orbits of spinning particles around a Kerr black hole \citep{Skoupy:2021b}.

A post-adiabatic term is of the order of radians and, thus, cannot be neglected. Hence, since the spin of the secondary contributes to the post-adiabatic term, we have to take it into account. In the case of compact objects, like black holes and neutron stars, a pole-dipole approximation is considered to be sufficient, and all the higher multipoles of the body can be ignored. The scalars describing a pole-dipole secondary are its mass $\mu$ and the measure of its spin $S$. In the EMRI framework, instead of $S$ we can gain more insight about the contribution of the secondary spin by defining its dimensionless counterpart $\sigma=S/(\mu M)$. For example, if we consider the secondary black hole as an extreme Kerr black hole, we have that $S=\mu^2$ leading to $\sigma=q$, which suggests that $\sigma$ is of the order of the mass ratio, i.e. $\sigma \lesssim q \ll 1$. This fact, actually, allows us to ignore all the terms with higher powers in $\sigma$ and focus on the linearized in spin contributions to the inspiral\footnote{This reasoning holds away from the resonances, since the resonances are governed by the $\order{S^2}$ \cite{Zelenka20}, which implies a contribution to the phase of order of radians.}. Hence, this work focuses on the influence of the secondary spin on the evolution of an inspiral moving on the equatorial plane of a Kerr black hole, when the calculations are restricted to the linear order in spin. Having confined our study on the equatorial plane of a Kerr black hole allows us to parametrize the orbital evolution by the energy $E$ and the $z$-component of the angular momentum $J_z$ of the system. The energy and the angular momentum fluxes, which reach infinity and the horizon, were already derived in \citep{Skoupy:2021b}. In this work we linearize these fluxes to calculate the adiabatic inspiral and the linear in spin part of the GW phase, i.e. the phase shift between the adiabatic inspiral of a spinning secondary and a non-spinning secondary. In particular, this phase shift $\delta\Phi_\mu$ can be found by linearizing in spin of the phase, i.e.
\begin{equation}
    \Phi_\mu = \frac{1}{q} \Phi^{(0)}_\mu + \frac{\sigma}{q} \delta \Phi_\mu + \order{\sigma^2/q} \; .
\end{equation}
Note that in this work we neglect the other post-adiabatic terms, the evolution of the primary mass and its spin due to the absorption of the GWs through the horizon and as well as the evolution of the spin magnitude $\sigma$. 

The rest of this paper is organized as follows. Sec.~\ref{sec:spinningParticles} describes the dynamics of a spinning body in a Kerr spacetime and introduces the orbital variables linearized in the spin of the secondary. Sec.~\ref{sec:GWFluxes} focuses on GW fluxes from spinning bodies moving on eccentric equatorial orbits around a Kerr black hole with these fluxes linearized in spin. Sec.~\ref{sec:evolution} presents the equations driving the adiabatic evolution of the orbital parameters and the phases. By linearization in spin this section provides the equations governing the phase shifts. Sec.~\ref{sec:implementationResults} first discusses the numerical methods and then provides the respective results. Finally, Sec.~\ref{sec:Concl} summarizes the main findings of our work. 

\textit{Notation:} In this work we use geometrized units where $c=G=1$. A partial derivative is denoted with a comma as $U_{\mu,\nu} = \partial_\nu U_\mu$ whereas a covariant derivative is denoted by a semicolon as $U_{\mu;\nu} = \nabla_\nu U_\mu$. The Riemann tensor is defined as $R^\mu{}_{\nu\kappa\lambda} = \Gamma^{\mu}{}_{\nu\lambda,\kappa} - \Gamma^{\mu}{}_{\nu\kappa,\lambda} + \Gamma^{\mu}{}_{\rho\kappa} \Gamma^{\rho}{}_{\nu\lambda} - \Gamma^{\mu}{}_{\rho\lambda} \Gamma^{\rho}{}_{\nu\kappa}$ and the signature of the metric is $(-,+,+,+)$. For convenience we use some quantities in their dimensionless form, which is denoted by a hat. A list with these quantities and their dimensionless counterparts can be found in Appendix~\ref{app:dimensionless}.

\section{Motion of a spinning particle}
\label{sec:spinningParticles}

Following Mathisson's gravitational skeleton approach \cite{Mathisson:1937zz,Mathisson:2010} and truncating the expansion, the stress-energy tensor of a spinning test body in a curved spacetime can be written as
\begin{equation} \label{eq:Tmunu}
    T^{\mu\nu} = \frac{1}{\sqrt{-g}} \left( \frac{P^{(\mu} v^{\nu)}}{v^t} \delta^3 - \nabla_\alpha \left( \frac{S^{\alpha(\mu} v^{\nu)}}{v^t} \delta^3 \right) \right)\, ,
\end{equation}
where $P^\mu$ is the four momentum, $v^\mu = \dv*{x^\mu}{\tau}$ is the four velocity, $S^{\alpha\beta}$ is the spin tensor, $\delta^3 \equiv \delta^3(x^i-x_p^i(t))$ is Dirac delta function located at the particle position $x_p^i(t)$ parametrized by the coordinate time $t$ and $g$ is the determinant of the metric. In this so called pole-dipole approximation the stress-energy tensor consists of a monopole (first term) and a dipole (second term). 

Applying the stress-energy conservation law $T^{\mu\nu}{}_{;\nu}=0$ on the stress-energy tensor \eqref{eq:Tmunu} the Mathisson-Papapetrou-Dixon (MPD) equations \citep{Mathisson:2010,Papapetrou:1951pa,Dixon:1964}  
\begin{subequations}
 \label{eq:MPEQs}
\begin{align} 
    \frac{{\rm D}P^\mu}{\rmd \tau} &= - \dfrac{1}{2} \; {R^\mu}_{\nu\rho\sigma} \; v^\nu \; S^{\rho\sigma} \; , \\ 
    \frac{{\rm D}S^{\mu\nu}}{\rmd \tau} & = P^\mu v^\nu - P^\nu v^\mu 
\end{align}
\end{subequations}
can be derived, where $R^{\mu}{}_{\nu\rho\sigma}$ is the Riemann tensor and $\tau$ is the proper time. 

The MPD system of equations is underdetermined, because for the 14 independent components $(x^\mu, P^\mu, S^{\mu\nu})$ only 10 independent equations are available. This ambiguity is related to the freedom we have to choose the centre of mass of the spinning body.  Thus, additional conditions must be imposed to fix the centre of mass and close the system. One such condition is the Tulczyjew-Dixon spin supplementary condition (TD SSC) \cite{tulczyjew1959motion,Dixon:1970zza}
\begin{align}
    S^{\mu\nu} P_\mu = 0  \; ,
\end{align}
which introduces three independent constraints to the system. The fourth constraint comes from the fact we have chosen the proper time as the evolution parameter in Eq.~\eqref{eq:MPEQs} and, hence,
\begin{align}\label{eq:4velCon}
 v^\mu v_\mu=-1 .
\end{align}
Note that in order to follow the evolution of the body, we actually track the worldline along the centre of the mass, which is the reason why a spinning body is often called a spinning particle. We will use both terms interchangeably throughout the rest of the paper.

Under the TD SSC the mass of the spinning particle with respect to the four-momentum
\begin{equation}
    \mu = \sqrt{-P^\mu P_\mu}
\end{equation}
and the magnitude of the particle's spin
\begin{equation}
    S = \sqrt{\frac{S^{\mu\nu}S_{\mu\nu}}{2}}
\end{equation}
are conserved along the trajectory. Often it is convenient to use the dimensionless spin parameter $\sigma$
\begin{equation}
    \sigma = \frac{S}{\mu M} \, ,
\end{equation}
instead of the spin magnitude $S$ and the spin four-vector 
\begin{align}
\label{eq:SpinVect}
 S_\mu = -\frac{1}{2} \epsilon_{\mu\nu\rho\sigma}
          \, u^\nu \, S^{\rho\sigma} \; 
\end{align}
instead the spin tensor, where $u^\mu = P^\mu/\mu$. It can be checked then, that the spin magnitude can be expressed as $S=\sqrt{S^\mu S_\mu}$.

Thanks to the SSC it is possible to derive a relation between the four momentum and the four velocity \citep{Ehlers1977}
\begin{align}
 \label{eq:v_p_TUL}
 v^\mu = \frac{\textsf{m}}{\mu} \left(
          u^\mu + 
          \frac{ 2 \; S^{\mu\nu} R_{\nu\rho\kappa\lambda} u^\rho S^{\kappa\lambda}}
          {4 \mu^2 + R_{\alpha\beta\gamma\delta} S^{\alpha\beta} S^{\gamma\delta} }
          \right)  \; ,
\end{align}
where $\textsf{m} \equiv -p^\mu v_\mu$ is the rest mass with respect to the four-velocity. The value of this mass is not conserved under TD SSC, however, it is constraint by Eq.~\eqref{eq:4velCon}.

\subsection{Motion on a Kerr background} 

We are interested in the motion of a spinning particle in Kerr spacetime background. This spacetime describes a spinning black hole at vacuum. The nonzero components of the Kerr metric in Boyer-Lindquist (BL) coordinates 
\begin{multline}
    \rmd s^2 = g_{tt}~\rmd t^2+2~g_{t\phi}~\rmd t~\rmd \phi + g_{\phi\phi}~\rmd \phi^2 \\
       + g_{rr}~\rmd r^2+g_{\theta\theta}~\rmd \theta^2  \label{eq:LinEl}
\end{multline}
read
 \begin{eqnarray}
   g_{tt} &=&-\left(1-\frac{2 M r}{\Sigma}\right) \; ,\nonumber\\ 
   g_{t\phi} &=& -\frac{2 a M r \sin^2{\theta}}{\Sigma} \; ,\nonumber\\
   g_{\phi\phi} &=& \frac{(\varpi^4-a^2\Delta \sin^2\theta) \sin^2{\theta}}{\Sigma} \; , \label{eq:KerrMetric}\\
   g_{rr} &=& \frac{\Sigma}{\Delta} \; ,\nonumber\\
   g_{\theta\theta} &=& \Sigma \nonumber
 \end{eqnarray} 
with
 \begin{eqnarray}
  \Sigma &=& r^2+ a^2 \cos^2{\theta} \; ,\nonumber\\
  \Delta &=& \varpi^2-2 M r \; ,\nonumber \\ 
  \varpi^2 &=& r^2+a^2 \; , \label{eq:Kerrfunc} 
 \end{eqnarray}
where $M$ is the mass of the black hole and $a$ is the Kerr parameter. 

The outer horizon of a Kerr black hole is located at $r_+ = M + \sqrt{M^2-a^2}$, and the spacetime is equipped with two killing vectors, one time-like $\xi^\mu_{(t)} = \delta^\mu_t$ and one space-like $\xi^\mu_{(\phi)} = \delta^\mu_\phi$. The existence of these Killing vectors provides the conservation of two additional quantities, namely of the energy measured at infinity
\begin{align}\label{eq:EnCons}
    E &= -P_t+\frac12g_{t\mu,\nu}S^{\mu\nu}
\end{align}
and of the total angular momentum projected onto the symmetry axis of the black hole measured at infinity
\begin{align}\label{eq:JzCons}
    J_z &= P_\phi-\frac12g_{\phi\mu,\nu}S^{\mu\nu} \, .
\end{align}

\subsection{Equatorial motion}

In our work we focus on the equatorial motion, hence $v^\theta=0$. It can be shown that in this case the particle stays in the equatorial plane \citep{Skoupy:2021b} and it holds that $p^\theta=0$ and 
\begin{equation}
    S_\mu= - r S \,\delta_\mu^\theta \,.
\end{equation}
Bounded equatorial orbits can be characterized by their semi-latus rectum $p$ and their eccentricity $e$, which are defined as
\begin{equation}
    p = \frac{2 \hr_1 \hr_2}{\hr_1 + \hr_2} \; , \qquad e = \frac{\hr_2 - \hr_1}{\hr_1 + \hr_2} \, ,
\end{equation}
where $\hr_1$ is the pericenter and $\hr_2$ is the apocenter. For the orbital description, we introduce dimensionless counterparts of the involved quantities (for details see Table~\ref{tab:dimensionless}). 

The radial coordinate of the particle periodically oscillates between $\hr_1$ and $\hr_2$. Because of this fact, we can change the parametrization of the trajectory from proper time $\tau$ to the angle-like relativistic anomaly $\chi$ defined as
\begin{equation}
    \hr = \frac{p}{1+e\cos(\chi+\chi_0)} \, ,
\end{equation}
where $\chi_0$ determines the initial radial position. For $\chi+\chi_0=0$ and $2\pi$ the particle is at the pericenter and for $\chi+\chi_0=\pi$ the particle is at the apocenter. The equations of motion for $t$ and $\phi$ in this so called Darwin parametrization then read
\begin{subequations}\label{eq:EOM_tphi}
\begin{align}
    \frac{\rmd \hat{t}}{\rmd\chi} &= V^t\left(\frac{p}{1+e\cos(\chi+\chi_0)}\right) \sqrt{\frac{1-e^2}{p^2 J(\chi+\chi_0)}} \, , \label{eq:EOM_t2} \\
    \frac{\rmd\phi}{\rmd\chi} &= V^\phi\left(\frac{p}{1+e\cos(\chi+\chi_0)}\right) \sqrt{\frac{1-e^2}{p^2 J(\chi+\chi_0)}} \, , \label{eq:EOM_phi2}
\end{align}
\end{subequations}
where the functions $V^t$, $V^\phi$ and $J$ can be found in Appendix~\ref{app:trajectory}. 

By integrating over $\chi$, the functions $\hat{t}(\chi)$ and $\phi(\chi)$ read
\begin{subequations}\label{eq:tchi_phichi}
\begin{align}
    \hat{t}(\chi) = \int_0^\chi \dv{\hat{t}}{\chi}\qty(\chi') \dd \chi' \; , \label{eq:tchi} \\
    \phi(\chi) = \phi_0 + \int_0^\chi \dv{\phi}{\chi}\qty(\chi') \dd \chi' \; , 
\end{align}
\end{subequations}
where we set the initial time $t(0)=0$.

Since, it is possible to express the energy and the angular momentum as $\hat{E}(p,e,\sigma)$, $\hat{J}_z(p,e,\sigma)$, i.e. as functions of $p$, $e$ and $\sigma$~\footnote{They also depend on the  Kerr parameter $a$, but we will treat it only as a parameter.} \citep{Skoupy:2021b} (see Appendix \ref{app:trajectory}), to uniquely identify a trajectory one needs four parameters $p,e,\chi_0,\phi_0$. However, many quantities are independent of the initial angles $\chi_0, \phi_0$. Therefore, we can define a fiducial trajectory with $\chi_0=0$ and $\phi_0=0$. The coordinates of this trajectory as well as all the quantities calculated from it are denoted with a check-mark as $\check{\hat{t}}(\chi)$, $\check{\hat{r}}(\chi)$, $\check{\phi}(\chi)$. After the 
substitution $\chi = v-\chi_0$, Eq. \eqref{eq:tchi} can be written as
\begin{equation}
    \hat{t}(\chi) = \int_{\chi_0}^{\chi+\chi_0} \dv{\check{\hat{t}}}{\chi}\qty(v) \dd v = \check{\hat{t}}(\chi+\chi_0) - \check{\hat{t}}(\chi_0) \; ,
\end{equation}
where $\dv*{\check{\hat{t}}}{\chi}$ comes from Eq. \eqref{eq:EOM_t2} when $\chi_0=0$. Analogous relation holds for $\phi(\chi)$ and, therefore, a general trajectory can be expressed using a fiducial trajectory as
\begin{subequations}\label{eq:GeneralFiducial}
\begin{align}
    \hat{t}(\chi) &= \check{\hat{t}}(\chi+\chi_0) - \check{\hat{t}}(\chi_0) \; , \\
    \hat{r}(\chi) &= \check{\hat{r}}(\chi+\chi_0) \; , \\
    \phi(\chi) &= \phi_0 +  \check{\phi}(\chi+\chi_0) - \check{\phi}(\chi_0) \; .
\end{align}
\end{subequations}
Trajectory-dependent quantities such as the frequencies or the GW fluxes, which are independent of $\chi_0$ and $\phi_0$, can be calculated using the fiducial trajectory. 

The radial period, i.e. the time between two successive passages through the pericenter can be expressed as
\begin{align}
    \hat{T}_r &= \frac{\sqrt{1-e^2}}{p} \int_0^{2\pi} V^t\left(\frac{p}{1+e\cos\chi}\right) \frac{1}{\sqrt{J(\chi)}} \dd \chi  \nonumber \\&= 2\frac{\sqrt{1-e^2}}{p} \int_0^\pi V^t\left(\frac{p}{1+e\cos\chi}\right) \frac{1}{\sqrt{J(\chi)}} \dd\chi \; ,
\end{align}
where we can integrate from 0 to $\pi$ because the integrand is even around $\pi$. Similarly, the accumulated phase of the azimuthal coordinate can be written as
\begin{equation}
    \Delta\phi = 2\frac{\sqrt{1-e^2}}{p} \int_0^\pi V^t\left(\frac{p}{1+e\cos\chi}\right) \frac{1}{\sqrt{J(\chi)}} \dd\chi \; .
\end{equation}

The frequencies with respect to the BL time can be then calculated as
\begin{subequations}\label{eq:Omegai}
\begin{align}
    \hat{\Omega}_r &= \frac{2\pi}{\hat{T}_r} \; , \label{eq:Omega_r}\\
    \hat{\Omega}_\phi &= \frac{\Delta\phi}{\hat{T}_r} \; . \label{eq:Omega_phi}
\end{align}
\end{subequations}

\subsection{Linearization in the secondary spin}

Due to the fact that the dimensionless spin $\sigma$ is of the same order as the mass ratio $q$, i.e. $\sigma \ll 1$, it is reasonable to linearize the expressions for the frequencies \eqref{eq:Omegai} in $\sigma$ to obtain
\begin{subequations}
\begin{equation}
    \hat{\Omega}_i(p,e,\sigma) = \hat{\Omega}_{i}^{\rm (g)}(p,e) + \sigma\, \delta\hat{\Omega}_i(p,e) + \order{\sigma^2} \; ,
\end{equation}
where $i=r,\phi$ and
\begin{align}
    \hat{\Omega}_{i}^{\rm (g)}(p,e) &= \hat{\Omega}(p,e,\sigma=0) \; , \\
    \delta\hat{\Omega}_i(p,e) &= \qty(\pdv{\hat{\Omega}_i}{\sigma})_{\sigma=0} \; .
\end{align}
\end{subequations}
Note that the index (g) in the above quantities refers to a geodesic orbit, i.e. for $\sigma=0$. 

However, for the calculation of GW fluxes it is convenient to linearize the quantities, such as energy and angular momentum fluxes, with respect to a reference geodesic with the same orbital frequencies (see Section \ref{sec:fluxeslin}). In other words, we must linearize the functions parametrized by the frequencies, i.e. $f(p(\hat{\Omega}_i,\sigma),e(\hat{\Omega}_i,\sigma),\sigma)$. For this, one must find the linear part of the functions
\begin{subequations} \label{eq:lin_p_e}
\begin{align}
    p(\hat{\Omega}_i,\sigma) &= p^{\rm (g)}(\hat{\Omega}_i) + \sigma\,\delta p(\hat{\Omega}_i) + \order{\sigma^2} \; , \\
    e(\hat{\Omega}_i,\sigma) &= e^{\rm (g)}(\hat{\Omega}_i) + \sigma\,\delta e(\hat{\Omega}_i) + \order{\sigma^2}\; ,
\end{align}
\end{subequations}
where $\delta p$ and $\delta e$ correspond to the change of the orbital parameters after a geodesic with frequencies $\hat{\Omega}_i$ is perturbed by a secondary spin $\sigma$ while keeping the frequencies same. Because the relations $p(\Omega_i,\sigma)$ and $e(\Omega_i,\sigma)$ are not known, we cannot simply take the derivative of $p(\Omega_i,\sigma)$ and $e(\Omega_i,\sigma)$ with respect to $\sigma$ to find $\delta p(\Omega_i)$ and $\delta e(\Omega_i)$, instead we have to use the derivatives of the implicit functions
\begin{subequations}
\begin{align}
   \hat{\Omega}_r &= \hat{\Omega}_r(p(\hat{\Omega}_r,\hat{\Omega}_\phi,\sigma), e(\hat{\Omega}_r,\hat{\Omega}_\phi,\sigma), \sigma) \; , \\
    \hat{\Omega}_\phi &= \hat{\Omega}_\phi(p(\hat{\Omega}_r,\hat{\Omega}_\phi,\sigma), e(\hat{\Omega}_r,\hat{\Omega}_\phi,\sigma), \sigma)
\end{align}
\end{subequations}
with respect to $\sigma$ to find them. In these functions the lhs is constant and the rhs are functions defined in Eqs.~\eqref{eq:Omegai}. After differentiating them with respect to $\sigma$, substituting $\sigma=0$ and solving for $\delta p = \pdv*{p}{\sigma}$ and $\delta e = \pdv*{e}{\sigma}$, we obtain
\begin{subequations}\label{eq:deltape}
\begin{align} \label{eq:deltap}
  \delta p &=   \frac{ \displaystyle  \pdv{\hat{\Omega}_{\phi}^{\rm (g)}}{e} \delta\hat{\Omega}_r-\pdv{\hat{\Omega}_{r}^{\rm (g)}}{e} \delta\hat{\Omega}_\phi}{\abs{J_{(\hat{\Omega}_i)}}} \; , \\
    \delta e &= \frac{ \displaystyle  -\pdv{\hat{\Omega}_{\phi}^{\rm (g)}}{p} \delta\hat{\Omega}_r+\pdv{\hat{\Omega}_{r}^{\rm (g)}}{p} \delta\hat{\Omega}_\phi}{\abs{J_{(\hat{\Omega}_i)}}} \; , \label{eq:deltae}
\end{align}
\end{subequations}
where all the derivatives are evaluated at $\sigma=0$ and the determinant of the Jacobian matrix is
\begin{equation} \label{eq:JacOmega_i}
    \abs{J_{(\hat{\Omega}_i)}} = \pdv{\hat{\Omega}_{r}^{\rm (g)}}{p} \pdv{\hat{\Omega}_{\phi}^{\rm (g)}}{e} -  \pdv{\hat{\Omega}_{r}^{\rm (g)}}{e} \pdv{\hat{\Omega}_{\phi}^{\rm (g)}}{p} \; .
\end{equation} 
Since $\delta p(p,e)$ and $\delta e(p,e)$ were derived through the above procedure, they are functions of $p$ and $e$. Actually, they can be interpreted as shifts of $p$ and $e$ when a geodesic originally with semi-latus rectum $p$ and eccentricity $e$ is perturbed by a spin $\sigma$, while keeping the frequencies constant.

As was proven in \citep{Barack:2011} for the Schwarzschild spacetime and in \citep{Warburton:2013} for the Kerr spacetime, bound geodesics cannot be uniquely parametrized by the frequencies $\Omega_{i}^{\rm (g)}$ and there exist a region of the parameter space near the separatrix with pairs of orbits with identical frequencies $\Omega_{r{\rm}}$ and $\Omega_{\phi{\rm}}$. This implies that there exists a curve in the $p-e$ plane sepatating these pairs, on which the determinant \eqref{eq:JacOmega_i} is zero. Therefore quantities linearized with respect to a geodesic with the same frequencies cannot be calculated on this curve.

The constants of motion $\hat{E}$ and $\hat{J}_z$ from Eqs.~\eqref{eq:energy2} and \eqref{eq:angmom2} are functions of $p,e$ and $\sigma$, hence the linear part in $\sigma$ with respect to a geodesic with the same frequencies can be found using the chain rule as
\begin{subequations} \label{eq:deltaE_deltaJz}
\begin{align}
    \eval{\delta \hat{E}}_{\Omega_i} &= \eval{\pdv{\hat{E}}{\sigma}}_{\sigma=0} + \pdv{\hat{E}^{\rm (g)}}{p} \delta p + \pdv{\hat{E}^{\rm (g)}}{e} \delta e \; , \\
    \eval{\delta \hat{J}_z}_{\Omega_i} &= \eval{\pdv{\hat{J}_z}{\sigma}}_{\sigma=0} + \pdv{\hat{J}_{z}^{\rm (g)}}{p} \delta p + \pdv{\hat{J}_{z}^{\rm (g)}}{e} \delta e \; ,
\end{align}
\end{subequations}
where $\delta p$ and $\delta e$ come from Eqs.~\eqref{eq:deltape} and the subscript $\Omega_i$ denotes that the quantity is linearized with respect to a geodesic with the same frequencies. We have, thus, introduced the operator $\eval{\delta f}_{\Omega_i}$ acting on a function $f(p,e,\sigma)$ as
\begin{equation}
    \eval{\delta f}_{\Omega_i} = \eval{\pdv{f}{\sigma}}_{\sigma=0} + \pdv{f^{\rm (g)}}{p} \delta p + \pdv{f^{\rm (g)}}{e} \delta e \, .
\end{equation}

\begin{figure}
  \centering  
  \includegraphics[width=0.48\textwidth]{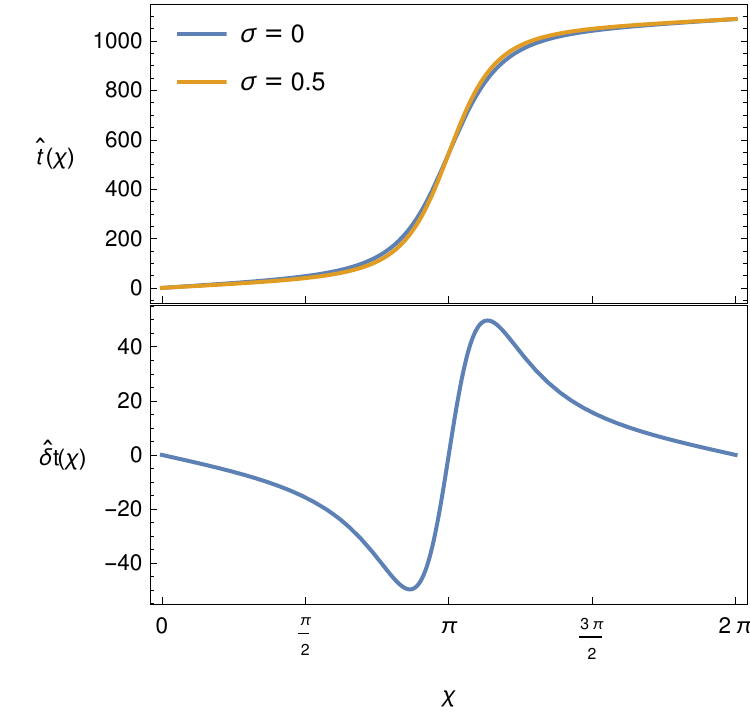}
  \caption{Top: The evolution of $\hat{t}^{\rm (g)}(\chi)$ for a geodesic orbit with $\hat{\Omega}_r = 0.00577033$, $\hat{\Omega}_\phi = 0.00942436$, which corresponds to $p^{\rm (g)}=10$, $e^{\rm (g)}=0.8$, and $t(\chi)$ for a trajectory of a spinning particle with $\sigma=0.5$ and the same frequencies as the geodesic orbit, which corresponds to $p=8.6538$, $e=0.831688$. Bottom: difference $\delta\hat{t}(\chi) = (\hat{t}(\chi)-\hat{t}^{\rm (g)}(\chi))/\sigma$. We can see that if the initial difference is $\hat{t}(0)-\hat{t}^{\rm (g)}(0)=0$, then at the end of the period $\hat{t}(2\pi)-\hat{t}^{\rm (g)}(2\pi)=0$ as well. The spin value has been chosen to be unphysically large to make the difference visible.}
  \label{fig:t_chi}
\end{figure}

Using the above linearized quantities, the coordinate functions $(\hat{t}(\chi),\hat{r}(\chi),\phi(\chi))$ can be linearized as well. When an equatorial geodesic parametrized by $\chi$ with frequencies $\hat{\Omega}_i$ is perturbed by a spin $\sigma$, the change of the coordinate time and the azimuthal coordinate can be described as
\begin{subequations}
\begin{align}
    \hat{t}(\chi) &= \hat{t}^{\rm (g)}(\chi) + \sigma\,\eval{\delta \hat{t}}_{\Omega_i}(\chi) + \order{\sigma^2} \; , \\
    \phi(\chi) &= \phi^{\rm (g)}(t) + \sigma\,\eval{\delta \phi}_{\Omega_i}(\chi) + \order{\sigma^2} \; ,
\end{align}
\end{subequations}
where $\hat{t}^{\rm (g)}(\chi),\phi^{\rm (g)}(\chi)$ are calculated from Eqs.~\eqref{eq:EOM_tphi} for $\sigma=0$ and equations for $\eval{\delta \hat{t}}_{\Omega_i}(\chi), \eval{\delta \phi}_{\Omega_i}(\chi)$ are derived by linearizing Eqs.~\eqref{eq:EOM_tphi} in $\sigma$ with respect to a geodesic with the same frequencies, i.e.
\begin{subequations} \label{eq:delta_t_phi}
\begin{align}
    \dv{\delta \hat{t}}{\chi} &= \eval{\pdv{\sigma}(\dv{\hat{t}}{\chi})}_{\sigma=0} + \pdv{p}(\dv{\hat{t}^{\rm (g)}}{\chi}) \delta p + \pdv{e}(\dv{\hat{t}^{\rm (g)}}{\chi}) \delta e \; , \\
    \dv{\delta \phi}{\chi} &= \eval{\pdv{\sigma}(\dv{\phi}{\chi})}_{\sigma=0} + \pdv{p}(\dv{\phi^{\rm (g)}}{\chi}) \delta p + \pdv{e}(\dv{\phi^{\rm (g)}}{\chi}) \delta e \; .
\end{align}
For the fiducial trajectory the initial conditions can be chosen such that the linear corrections $\delta t$, $\delta \phi$ are zero at the pericenter, namely $\delta\check{\hat{t}}(0)=0=\delta\check{\phi}(0)$. Thanks to the frequency matching, it holds $\delta\check{\hat{t}}(2\pi) = 0 = \delta\check{\phi}(2\pi)$, because the radial period and accumulated phase in $\phi$ are the same for both the perturbed and the unperturbed trajectory. This can be seen in Fig.~\ref{fig:t_chi}, where we plot the evolution of $\hat{t}(\chi)$ for a geodesic orbit ($\sigma=0$) with $p=10$, $e=0.8$ and for a trajectory of a spinning particle with $\sigma=0.5$, which frequencies were matched to the same as the frequencies of the geodesic orbit.

The linear correction to the radial coordinate can be calculated as
\begin{align}
 \delta\hr(\chi) &= \pdv{\hr}{p} \delta p + \pdv{\hr}{e} \delta e  \nonumber \\
 &= \frac{\delta p}{1+e \cos\chi} - \frac{p\,\delta e \cos\chi}{(1+e\cos\chi)^2} \; .
\end{align}
\end{subequations}

\section{Gravitational-wave fluxes}
\label{sec:GWFluxes}

For the calculation of the GW fluxes we use Teukolsky formalism where the GWs are treated as perturbations of the background spacetime. To obtain the GW fluxes to infinity and to the horizon we calculate perturbation of the Weyl curvature scalar 
\begin{equation}
    \Psi_4 = -C_{\alpha\beta\gamma\delta} n^\alpha \mbar^\beta n^\gamma \mbar^\delta \, ,
\end{equation}
where $C_{\alpha\beta\gamma\delta}$ is the Weyl tensor and
\begin{align}
    n^\mu &= \frac{1}{2\Sigma}\left( \varpi^2 , - \Delta , 0, a \right) \; , \\
    \mbar^\mu  &= -\frac{1}{\sqrt{2}\zeta} \left( ia\sin\theta , 0,  -1,   i\csc\theta \right)
\end{align}
are two legs of the Kinnersley null tetrad with $$\zeta=r-i a \cos\theta.$$ 

The Weyl scalar is related to the gravitational radiation at infinity as
\begin{equation} \label{eq:Psi4}
    \Psi_4(r \rightarrow \infty) = \frac{1}{2} \dv[2]{h}{t} \; ,
\end{equation}
where $h = h_+ - i h_\times$ is the strain, which is defined as $h_{\mu\nu} = h_+ e_{\mu\nu}^+ + h_\times e_{\mu\nu}^\times$ with the metric perturbation $h_{\mu\nu}$ and polarization tensors $e_{\mu\nu}^{{+},\times}$. The Weil scalar $\Psi_4$ encodes the gravitational radiation emitted to infinity; however, by using the Teukolsky-Starobinsky identities, it is possible to infer from $\Psi_4$ the fluxes at the horizon as well.

Teukolsky in \citep{Teukolsky:1973ha} introduced the master equation for the field in the form\footnote{In this section the coordinates $(t,r,\theta,\phi)$ denote an event in the spacetime in which the field is measured, while the trajectory of the particle is denoted by $(t_{\rm p},r_{\rm p},\theta_{\rm p},\phi_{\rm p})$.}
\begin{equation} \label{eq:teuk}
    {}_s\mathcal{O} \, {}_s\psi(t,r,\theta,\phi) = 4\pi \Sigma T \; ,
\end{equation}
where $_s\mathcal{O}$ is a second order partial differential operator and $T$ is a source term calculated as certain differential operator acting on projections of the stress energy tensor (the interested reader is referred to \citep{Teukolsky:1973ha} for more details). In the case of GWs, the calculated quantity from Eq.~\eqref{eq:teuk} is $_{-2}\psi = \zeta^4 \Psi_4$.

In this paper we use frequency domain solutions of the Teukolsky equation (TE), for which the field is written using Fourier modes
\begin{equation} \label{eq:psi_fourier}
    {}_{-2}\psi = \sum_{l,m}^{\infty} \frac{1}{2\pi} \int_{-\infty}^{\infty} \rmd \omega\, \psi_{lm\omega}(r) {}_{-2}S_{lm}^{a\omega}(\theta) e^{-i\omega t + i m \phi} \; .
\end{equation}
Having done that, Eq. \eqref{eq:teuk} can be separated into two ordinary differential equations: one for the radial part $\psi_{lm\omega}(r)$ and one for the angular part ${}_{-2}S_{lm}^{a\omega}(\theta)$, which is called spin-weighted spheroidal harmonic.

The asymptotic behavior of the radial part at infinity and at the horizon can be written as \citep{Hughes:2021}
\begin{subequations}
\begin{align} \label{eq:psi_radial}
    \psi_{lm\omega}(r) &\approx C^+_{lm\omega} r^3 e^{i \omega r^\ast} \qquad r \rightarrow \infty \; , \\
     \psi_{lm\omega}(r) &\approx C^-_{lm\omega} \Delta e^{ - i k_{\mathcal{H}} r^\ast} \qquad r \rightarrow r_+ \; ,
\end{align}
\end{subequations}
where $k_{\mathcal{H}} = \omega - m\Omega_{\mathcal{H}}$ is the frequency at the horizon, $\Omega_{\mathcal{H}} = a/(2Mr_+)$ is the horizon's angular velocity and $r^\ast$ is the tortoise coordinate defined as $\dv*{r^\ast}{r}=\varpi^2/\Delta$. 

The amplitudes $C^\pm_{lm\omega}$ can be calculated using Green function formalism as
\begin{equation}
    C^\pm_{lm\omega} = \int_{-\infty}^{\infty} \rmd t\, e^{i\omega t - i m \phi_p(t)} I^\pm_{lm\omega}(r_{\rm p}(t),\theta_{\rm p}(t))
\end{equation}
with 
\begin{multline} \label{eq:Ipm}
    I^\pm_{lm\omega}(r,\theta) = \frac{1}{W} \left(A_0 - (A_1+B_1) \frac{\rmd}{\rmd r} \right. \\ \left. + (A_2+B_2) \frac{\rmd^2}{\rmd r^2} - B_3 \frac{\rmd^3}{\rmd r^3} \right) R^{\mp}_{lm\omega}(r) \; ,
\end{multline}
where $R^{\pm}_{lm\omega}(r)$ are homogeneous solutions of the radial equation satisfying boundary conditions at infinity ``+" or at the horizon ``-" respectively, $W$ is the invariant Wronskian and $A_i$, $B_i$ are functions of the orbital quantities. These quantities can be found in Appendix B of \citep{Skoupy:2021b}.

After we confine the particle trajectory into the equatorial plane, it can be shown that thanks to the periodicity of the radial motion the frequency spectrum is discrete and the amplitudes can be written as a sum over individual $n$-modes
\begin{equation}
    C^\pm_{lm\omega} = \sum_{n=-\infty}^\infty C^\pm_{lmn} \delta(\omega - \omega_{mn}) \label{eq:C_lmomega}
\end{equation}
with frequencies
\begin{equation}\label{eq:omega_mn}
    \omega_{mn} = m\Omega_\phi + n\Omega_r \; ,
\end{equation}
where $n$ is an integer.

After reparametrization of the orbit with $\chi$, the partial amplitudes can be caculated as
\begin{multline} \label{eq:Cpm_lmn_final}
    C^\pm_{lmn} = \Omega_r \int_0^\pi  \rmd \chi \sum_{D_r = \pm} \frac{\rmd t}{\rmd \chi} I^\pm_{lmn}(r_{\rm p}(\chi),\pi/2,D_r) \\ \times \exp(i D_r \varphi_{mn}(\chi) ) \; ,
\end{multline}
where $I^\pm_{lmn}=I^\pm_{lm\omega_{mn}}$, $\varphi_{mn}(\chi) = \omega_{mn} t_{\rm p}(\chi) - m \phi_{\rm p}(\chi)$ and $D_r$ is the sign of the radial velocity.

After Eqs. \eqref{eq:GeneralFiducial} are substituted into the above equation and the integration variable $\chi \rightarrow \chi - \chi_0$ is changed, the partial amplitudes from an equatorial orbit with $\chi_0\neq0,~\phi_0\neq0$ can be expressed using partial amplitudes from the fiducial trajectory $\check{C}^\pm_{lmn}$ and a phase factor as
\begin{equation}\label{eq:PartAmpRel}
    C^{\pm}_{lmn} = e^{i \xi_{mn}} \check{C}^{\pm}_{lmn} \; ,
\end{equation}
where the phase factor reads
\begin{equation}
    \xi_{mn} = -\omega_{mn} \check{t}_{\rm p}(\chi_0) + m \qty(\check{\phi}_{\rm p}(\chi_0) - \phi_0) \; .
\end{equation}
This factor agrees with Eq. (3.19) in \citep{Hughes:2021} for equatorial motion.

From Eqs. \eqref{eq:Psi4}, \eqref{eq:psi_fourier} and \eqref{eq:psi_radial} the strain at infinity can be expressed as
\begin{equation} \label{eq:strain}
    h = -\frac{2}{r} \sum_{lmn} \frac{C^+_{lmn}}{\omega_{mn}^2} S_{lm}^{a\omega_{mn}}(\theta) e^{-i\omega_{mn}(t-r^\ast) + im\phi} \; .
\end{equation}

The effective stress-energy of a GW can be reconstructed from the strain. From it, the orbit-averaged energy and angular momentum fluxes to the future null infinity $\mathcal{J}^+$ can be derived as
\begin{subequations}\label{eq:fluxes}
\begin{align}
    \left\langle\mathcal{F}^{E \mathcal{J}^+}\right\rangle &= \sum_{l=2}^\infty \sum_{m=-l}^l \sum_{n=-\infty}^\infty \frac{\left|\check{\hat{C}}^+_{lmn}\right|^2}{4\pi \hat{\omega}_{mn}^2} \; , \\
    \left\langle\mathcal{F}^{J_z \mathcal{J}^+}\right\rangle &= \sum_{l=2}^\infty \sum_{m=-l}^l \sum_{n=-\infty}^\infty \frac{m\left|\check{\hat{C}}^+_{lmn}\right|^2}{4\pi \hat{\omega}_{mn}^3} \; ,
\end{align}
where the brackets denote averaging over the radial period. Similar relations can be derived for the fluxes through the future horizon $\mathcal{H}^+$
\begin{align}
    \left\langle\mathcal{F}^{E \mathcal{H}^+}\right\rangle &= \sum_{l=2}^\infty \sum_{m=-l}^l \sum_{n=-\infty}^\infty \alpha_{lmn}\frac{\left|\check{\hat{C}}^-_{lmn}\right|^2}{4\pi \hat{\omega}_{mn}^2} \; ,\\
    \left\langle\mathcal{F}^{J_z \mathcal{H}^+}\right\rangle &= \sum_{l=2}^\infty \sum_{m=-l}^l \sum_{n=-\infty}^\infty \alpha_{lmn}\frac{m\left|\check{\hat{C}}^-_{lmn}\right|^2}{4\pi \hat{\omega}_{mn}^3} \; ,
\end{align}
\end{subequations}
where $\alpha_{lmn} = \alpha_{lm\omega_{mn}}$ can be found in \citep{Skoupy:2021b}. These fluxes are defined from the dimensionless quantities in accordance with \citep{Skoupy:2021b}. Note that thanks to the absolute value of the partial amplitudes in Eqs.~\eqref{eq:fluxes}, the phase correction in Eq.~\eqref{eq:PartAmpRel} is cancelled and, thus, the averaged fluxes can be computed from the fiducial trajectory.

\subsection{Linearization in the secondary spin}
\label{sec:fluxeslin}

The partial amplitudes $C^\pm_{lmn}$ calculated above depend on $p$, $e$ and $\sigma$, but since the formula contains the dependence on the frequencies $\Omega_i(p,e,\sigma)$, the partial amplitudes can be written as $C^\pm_{lmn}(p,e,\Omega_i(p,e,\sigma),\sigma)$. In this form, they can be linearized in $\sigma$ as
\begin{equation}
    C^\pm_{lmn}(p,e,\sigma) = C^{\rm (g)\pm}_{lmn}(p,e) + \sigma \eval{\delta C^\pm_{lmn}}_{p,e}(p,e) + \order{\sigma^2} \; ,
\end{equation}
where
\begin{equation} \label{eq:deltaCpm}
    \eval{\delta C^\pm_{lmn}}_{p,e}(p,e) = \eval{\pdv{C^\pm_{lmn}}{\sigma}}_{\sigma=0} + \pdv{C^{\rm (g)\pm}_{lmn}}{\Omega_i} \pdv{\Omega_i}{\sigma} \; ,
\end{equation}
in which we use the convention that all repeated indices are summed over and $|_{p,e}$ denotes that the quantity is calculated with respect to reference geodesic with fixed $p,e$. The partial amplitudes $C^\pm_{lmn}$ depend on the frequencies $\Omega_i$, since the functions giving it, like the homogeneous solutions $R^\pm_{lm\omega}(r)$ and $S_{lm}^{a\omega}(\theta)$, depend on $\omega_{mn}$. Thus, for the calculation of $\eval{\delta C^\pm_{lmn}}_{p,e}(p,e)$ one needs the derivatives of $R_{lm\omega}(r)$ and $S_{lm}^{a\omega}(\theta)$ with respect to $\omega$.  To find these derivatives, the radial and angular TE must be differentiated with respect to $\omega$ and then this system of equation must be solved. Rather than developing a code for finding these derivatives, we were able to achieve our goal by calculating slightly different quantities, for which the $\omega$ derivative of the homogeneous solutions is not needed. In this alternative procedure, we can use the TE solver implemented in the Black Hole Perturbation Toolkit. In particular, we calculate the linear part of the partial amplitudes with respect to a reference geodesic with the same frequencies. Formally, the dependence of the partial amplitudes on $\Omega_i$ can be written as $C^\pm_{lmn}(p(\Omega_i,\sigma),e(\Omega_i,\sigma),\Omega_i,\sigma)$ which can be linearized as
\begin{equation}
    C^\pm_{lmn}(\Omega_i,\sigma) = C^{\rm (g)\pm}_{lmn}(\Omega_i) + \sigma\, \eval{\delta C^\pm_{lmn}}_{\Omega_i}(\Omega_i) + \order{\sigma^2} \; ,
\end{equation}
where
\begin{equation} \label{eq:deltaCOmega_lm_general}
    \eval{\delta C^\pm_{lmn}}_{\Omega_i}(\Omega_i) = \eval{\pdv{C^\pm_{lmn}}{\sigma}}_{\sigma=0} + \pdv{C^{\rm (g)\pm}_{lmn}}{p} \delta p + \pdv{C^{\rm (g)\pm}_{lmn}}{e} \delta e 
\end{equation}
and $\delta p$ and $\delta e$ are defined in Eqs. \eqref{eq:deltape}. All the above derivatives are calculated for $\sigma=0$, i.e. for a geodesic, and we can use the fact that $\Omega_i = \Omega_{i}^{\rm (g)}(p,e)$ to obtain these linear parts as functions of $p$ and $e$.

The linearized expression for $\delta C^\pm_{lmn}(p,e)$ from Eqs. \eqref{eq:Cpm_lmn_final} and \eqref{eq:deltaCOmega_lm_general} reads
\begin{multline} \label{eq:deltaCpm_lmn}
    \delta C^\pm_{lmn} = \Omega_r \int_0^\pi \dd\chi \sum_{D_r} \exp(i D_r \varphi_{mn}(\chi) ) \\ \times \qty( \eval{\delta\tilde{I}^\pm_{lmn}}_{\Omega_i} + \tilde{I}^\pm_{lmn} i D_r \delta\varphi_{mn}(\chi) )
\end{multline}
where $\tilde{I}^\pm_{lmn} = \dv*{t}{\chi} I^\pm_{lmn}$ and 
\begin{equation}
\delta\varphi_{mn}(\chi) = \omega_{mn} \delta t(\chi) - m \delta \phi(\chi) \; .    
\end{equation}

From Eqs.~\eqref{eq:fluxes} we can find the linear in $\sigma$ part of the fluxes $\mathcal{F}(p,e,\sigma) = \mathcal{F}^{\rm (g)}(p,e) + \sigma\,\delta\mathcal{F}(p,e)+\order{\sigma^2}$ where $\mathcal{F}$ stands for $\mathcal{F}^{E \mathcal{J}^+}$, $\mathcal{F}^{E \mathcal{H}^+}$, $\mathcal{F}^{J_z \mathcal{J}^+}$ and $\mathcal{F}^{J_z \mathcal{H}^+}$. The result is
\begin{multline} \label{eq:deltaF}
    \delta \mathcal{F}^{\mathcal{C} \mathcal{I}^+,\mathcal{H}^+} = \sum_{lmn}\Big( \Re{\delta \hat{C}^\pm_{lmn}} \Re{\hat{C}^{\rm (g)\pm}_{lmn}} \\ + \Im{\delta \hat{C}^\pm_{lmn}} \Im{\hat{C}^{\rm (g)\pm}_{lmn}} \Big) \frac{\hat{\beta}^{\pm}_{\mathcal{C}lmn}}{2\pi\hat{\omega}_{mn}^3}
\end{multline}
where $\mathcal{C}$ stands for $E$ or $J_z$ and $\hat{\beta}^+_{Elmn} = \hat{\omega}_{mn}$, $\hat{\beta}^+_{J_z lmn} = m$, $\hat{\beta}^-_{E lmn} = \alpha_{lmn} \hat{\omega}_{mn}$, $\beta^-_{J_z lmn} = \alpha_{lmn} m$. All the linear parts above are with respect to geodesic with the same frequencies.

When the geodesic fluxes and their linear corrections are calculated on a grid in the $p-e$ plane, it is possible to find the linear part $\eval{\delta\mathcal{F}}_{p,e}$ from $\eval{\delta\mathcal{F}}_{\Omega_i}$ and the derivatives of $\mathcal{F}^{\rm (g)}$ with respect to $p$ and $e$. Namely, 
\begin{equation}\label{eq:dF_peFromdF_Omega}
    \eval{\delta\mathcal{F}}_{p,e} = \eval{\delta\mathcal{F}}_{\Omega_i} - \dv{\mathcal{F}^{\rm (g)}}{p} \delta p - \dv{\mathcal{F}^{\rm (g)}}{e} \delta e \; ,
\end{equation}
where $\eval{\delta\mathcal{F}}_{\Omega_i}$ is computed using Eq. \eqref{eq:deltaF}, $\delta p$ and $\delta e$ are from Eqs. \eqref{eq:deltape} and the derivatives with respect to $p$ and $e$ are understood as
\begin{equation}\label{eq:dFgdp}
    \dv{\mathcal{F}^{\rm (g)}}{p,e} = \pdv{\mathcal{F}^{\rm (g)}}{p,e} + \pdv{\mathcal{F}^{\rm (g)}}{\hat{\Omega}_i} \pdv{\hat{\Omega}_i}{p,e} \; ,
\end{equation}
although, in our scheme, they are directly calculated numerically on the grid in the $p-e$ plane. 

Let us now prove that Eq.~\eqref{eq:dF_peFromdF_Omega} holds. The linear part $\eval{\delta\mathcal{F}}_{\Omega_i}$ reads
\begin{equation}
    \eval{\delta\mathcal{F}}_{\Omega_i} = \eval{\pdv{\mathcal{F}}{\sigma}}_{\sigma=0} + \pdv{\mathcal{F}^{\rm (g)}}{p} \delta p + \pdv{\mathcal{F}^{\rm (g)}}{e} \delta e \; ,
\end{equation}
since $\hat{\Omega}_i(p,e,\sigma)$. Replacing the above along with the  total derivatives with respect to $p$ and $e$ (Eq.~\eqref{eq:dFgdp}) into Eq.~\eqref{eq:dF_peFromdF_Omega}  reduces Eq.~\eqref{eq:dF_peFromdF_Omega}  to
\begin{equation}
    \eval{\delta\mathcal{F}}_{p,e} = \eval{\pdv{\mathcal{F}}{\sigma}}_{\sigma=0}  - \pdv{\mathcal{F}^{\rm (g)}}{\hat{\Omega}_i} \qty( \pdv{\hat{\Omega}_{i}^{\rm (g)}}{p} \delta p + \pdv{\hat{\Omega}_{i}^{\rm (g)}}{e} \delta e )
\end{equation}
By substituting Eqs.~\eqref{eq:deltape} into that latter, it can be proven that the term in brackets equals to $-\pdv*{\hat{\Omega}_i}{\sigma}$ and we, thus, obtain 
\begin{equation}
    \eval{\delta\mathcal{F}}_{p,e} = \eval{\pdv{\mathcal{F}}{\sigma}}_{\sigma=0} + \pdv{\mathcal{F}^{\rm (g)}}{\hat{\Omega}_i} \pdv{\hat{\Omega}_i}{\sigma}
\end{equation}
which is the definition of $\eval{\delta \mathcal{F}}_{p,e}$ similar to Eq. \eqref{eq:deltaCpm}.

Note that though the linear part $\eval{\delta\mathcal{F}}_{\Omega_i}$ is singular for some points on the $p-e$ plane due to a vanishing $\abs{J_{(\Omega_i)}}$ (Eq.~\eqref{eq:JacOmega_i}), the linear part $\eval{\delta\mathcal{F}}_{p,e}$ is regular in the whole parameter space for which the semi-latus rectum $p$ is larger than the separatrix one $p_{\rm s}$. This is caused by the cancellation of the diverging terms in $\eval{\delta\mathcal{F}}_{\Omega_i}$, $\delta p$ and $\delta e$ in Eq.~\eqref{eq:dF_peFromdF_Omega}. However, due to numerical errors arising in the calculation of $\dv*{\mathcal{F}^{\rm (g)}}{p,e}$, the result is not reliable near these diverging points and the error may be high.

\section{Adiabatic evolution of the orbits}
\label{sec:evolution}

During an equatorial inspiral, the orbital parameters $p$ and $e$ are slowly evolving due to gravitational radiation reaction. Using the adiabatic approximation in the framework of the two timescale approximation, thanks to the balance law, the evolution of an inspiral can be calculated from the energy and angular momentum fluxes to infinity and to the horizon \citep{Akcay:2020}. In particular, the evolution of the constants of motion is related to the averaged fluxes as
\begin{subequations}
\begin{align}
    \left\langle\dot{\hat{E}}\right\rangle &\equiv \left\langle\dv{\hat{E}}{\hat{t}}\right\rangle = - q \qty(\left\langle\mathcal{F}^{E \mathcal{I}^+}\right\rangle + \left\langle\mathcal{F}^{E \mathcal{H}^+}\right\rangle ) \; , \\
    \left\langle\dot{\hat{J}}_z\right\rangle &\equiv \left\langle\dv{\hat{J}_z}{\hat{t}}\right\rangle = - q \qty(\left\langle\mathcal{F}^{J_z \mathcal{I}^+}\right\rangle + \left\langle\mathcal{F}^{J_z \mathcal{H}^+}\right\rangle ) \; ,
\end{align}
\end{subequations}

Using the chain rule, the derivatives of $E$ and $J_z$ can be calculated from the derivatives of $p$ and $e$ as
\begin{equation}
    \mqty( \displaystyle \dv{\hat{E}}{\hat{t}} \\ \displaystyle \dv{\hat{J}_z}{\hat{t}} ) = \mqty( \displaystyle \pdv{\hat{E}}{p} & \displaystyle \pdv{\hat{E}}{e} \\ \displaystyle \pdv{\hat{J}_z}{p} & \displaystyle \pdv{\hat{J}_z}{e} ) \mqty( \displaystyle \dv{p}{\hat{t}} \\ \displaystyle \dv{e}{\hat{t}} ) \, .
\end{equation}
By inverting the Jacobian matrix we obtain the equations for $\dot{p}$ and $\dot{e}$ in the form
\begin{subequations}\label{eq:dpdtdedt}
\begin{align}
    \dv{p}{\hat{t}} &=  \frac{ \displaystyle \pdv{\hat{J}_z}{e} \dot{\hat{E}} - \pdv{\hat{E}}{e} \dot{\hat{J}}_z}{\abs{J_{(\hat{E},\hat{J}_z)}}} \equiv \dot{p}(p(\hat{t}),e(\hat{t}),\sigma) \; , \\
    \dv{e}{\hat{t}} &=   \frac{\displaystyle -\pdv{\hat{J}_z}{p} \dot{\hat{E}} + \pdv{\hat{E}}{p} \dot{\hat{J}}_z}{\abs{{J}_{(\hat{E},\hat{J}_z)}}} \equiv \dot{e}(p(\hat{t}),e(\hat{t}),\sigma) \; ,
\end{align}
\end{subequations}
where we have omitted the angle brackets for simplicity and where the Jacobian determinant is
\begin{equation}
    \abs{J_{(\hat{E},\hat{J}_z)}} = \pdv{\hat{E}}{p} \pdv{\hat{J}_z}{e} - \pdv{\hat{E}}{e} \pdv{\hat{J}_z}{p} \; .
\end{equation}
Thanks to Eq.~\eqref{eq:dpdtdedt}, the evolution of $p$ and $e$ can be computed using the fluxes which, in fact, depend on $p$ and $e$.

Once we have the evolution of $p(\hat{t})$ and $e(\hat{t})$, the waveform at infinity can be computed from Eq.~\eqref{eq:strain} as \citep{Pound:2021}
\begin{equation}
    \hat{r} h(\hat{u}) = \sum_{lmn} \hat{A}_{lmn}(\hat{u}) S^{\ha\hat{\omega}_{lm}(p(\hat{u}),e(\hat{u}))}_{lm}(\theta) e^{-i \Phi_{mn}(\hat{u}) + i m \phi} \; ,
\end{equation}
where $\hat{u}=\hat{t}-\hr^\ast$ is the retarded coordinate and the amplitudes and phases respectively read
\begin{align}
    \hat{A}_{lmn}(\hat{u}) &= -2 q \frac{\hat{C}^+_{lmn}(p(\hat{u}),e(\hat{u}))}{\hat{\omega}_{mn}^2(p(\hat{u}),e(\hat{u}))} \; \label{eq:AmplPart} , \\
    \Phi_{mn}(\hat{u}) &= \int_0^{\hat{u}} \hat{\omega}_{lm}(p(\hat{u}'),e(\hat{u}')) \dd \hat{u}' \; \label{eq:PhasPart}.
\end{align}
From Eq. \eqref{eq:omega_mn} the phase can be written as $\Phi_{mn} = m\Phi_\phi + n\Phi_r$, where the particular phases
\begin{equation} \label{eq:Phase}
    \Phi_i(\hat{u}) = \int_0^{\hat{u}} \hat{\Omega}_i(p(\hat{u}'),e(\hat{u}')) \dd \hat{u}'
\end{equation}
can be calculated separately. The partial amplitudes $C^+_{lmn}(p(\hat{u}),e(\hat{u}))$ can be calculated from the fiducial partial amplitude $\check{C}^+_{lmn}$ and the phase factor $\xi_{mn}(p(\hat{u}),e(\hat{u}))$, which evolves over time. This correction changes slowly and remains at the order of unity \citep{Hughes:2021}.

Note that the above amplitudes~\eqref{eq:AmplPart} and phases~\eqref{eq:PhasPart} are part of the two-timescale expansion in the first-order perturbation theory \cite{Pound:2021}. However, with modifications, this scheme can be used even in the calculations of second-order perturbations \citep{Miller:2021}.

\subsection{Linearization in the secondary spin}
\label{sec:EvolLin}

The evolution equations \eqref{eq:dpdtdedt} of $p$ and $e$ depend on $p$, $e$ and $\sigma$. Therefore, the evolution can be linearized in $\sigma$ as
\begin{subequations}
\begin{align}
    p(\hat{t},\sigma) &= p^{\rm (g)}(\hat{t}) + \sigma\,\delta p(\hat{t}) + \order{\sigma^2} \\
    e(\hat{t},\sigma) &= e^{\rm (g)}(\hat{t}) + \sigma\,\delta e(\hat{t}) + \order{\sigma^2}
\end{align}
\end{subequations}
where $p^{\rm (g)}(\hat{t}), e^{\rm (g)}(\hat{t})$ describe inspirals with non-spinning secondary and $\delta p(\hat{t}), \delta e(\hat{t})$ are corrections to the evolution due to the secondary spin\footnote{Note that these quantities are different from the quantities in Eqs. \eqref{eq:deltape}, which denote the change in the orbital parameters when a geodesic is perturbed with secondary spin while keeping the frequencies constant.}.

Functions $p^{\rm (g)}(\hat{t}), e^{\rm (g)}(\hat{t})$ are calculated from Eqs. \eqref{eq:dpdtdedt} for $\sigma=0$
\begin{subequations}\label{eq:dp0e0dt}
\begin{align}
    \dv{p^{\rm (g)}}{\hat{t}} &= \dot{p}(p^{\rm (g)}(\hat{t}), e^{\rm (g)}(\hat{t}),0) \, , \\
    \dv{e^{\rm (g)}}{\hat{t}} &= \dot{e}(p^{\rm (g)}(\hat{t}), e^{\rm (g)}(\hat{t}),0) \, ,
\end{align}
\end{subequations}
and $\delta p(t), \delta e(t)$ are calculated from the linear part of Eqs. \eqref{eq:dpdtdedt}
\begin{subequations}\label{eq:ddeltapedt}
\begin{align}
    \dv{\delta p}{\hat{t}} &= \eval{\dv{\dot{p}}{\sigma}}_{\sigma=0} \equiv \delta\dot{p}(p^{\rm (g)}(\hat{t}),e^{\rm (g)}(\hat{t}),\delta p(\hat{t}), \delta e(\hat{t})) \; , \\
    \dv{\delta e}{\hat{t}} &= \eval{\dv{\dot{e}}{\sigma}}_{\sigma=0} \equiv \delta \dot{e}(p^{\rm (g)}(\hat{t}),e^{\rm (g)}(\hat{t}), \delta p(\hat{t}), \delta e(\hat{t})) \; ,
\end{align}
\end{subequations}
where the total derivatives are defined as
\begin{equation}
    \eval{\dv{f}{\sigma}}_{\sigma=0} = \eval{\pdv{f}{\sigma}}_{\sigma=0} + \pdv{f^{\rm (g)}}{p} \delta p + \pdv{f^{\rm (g)}}{e} \delta e \; .
\end{equation}
More explicit formulas can be found in Appendix \ref{app:linearized_inspiral}.

The linear parts of $\dot{\hat{E}}$ and $\dot{\hat{J}}_z$ in Eqs.~\eqref{eq:ddeltapedt} are calculated from the linearized fluxes with respect to geodesic with the same $p$ and $e$, i.e. from $\eval{\delta\mathcal{F}}_{p,e}$, which is computed from Eq.~\eqref{eq:dF_peFromdF_Omega}. This equation as well as Eqs.~\eqref{eq:ddeltapedt} contain derivatives of the geodesic fluxes $\mathcal{F}^{\rm (g)}$ with respect to $p$ and $e$ which must be calculated numerically.

 After we expand the phase in the secondary spin as
\begin{equation}
    \Phi_i(\hat{u},\sigma) = \Phi^{\rm (g)}_i(\hat{u}) + \sigma \delta\Phi_i(\hat{u}) + \order{\sigma^2} \; ,
\end{equation}
we get the leading adiabatic term $\Phi^{\rm (g)}_i$, which is $\order{q^{-1}}$, and the linear in spin term together with the spin value $\sigma \delta\Phi_i$, which is $\order{\sigma/q} = \order{1}$. Since for LISA data analysis the GW phase is needed with precision to fractions of radians, apart from the former dominant term, also the the latter term must be included. In this work, we call $\sigma\,\delta\Phi_i$ a \emph{phase shift}. The linear in spin term can be calculated by the linearization of Eq. \eqref{eq:Phase} as
\begin{equation}\label{eq:deltaPhi}
    \delta \Phi_i = \int_0^{\hat{u}} \qty( \eval{\pdv{\hat{\Omega}_i}{\sigma}}_{\sigma=0} + \pdv{\hat{\Omega}_{i}^{\rm (g)}}{p} \delta p(\hat{u}') + \pdv{\hat{\Omega}_{i}^{\rm (g)}}{e} \delta e(\hat{u}')) \dd \hat{u}' \, ,
\end{equation}
where the derivatives of $\Omega_{i}$ are evaluated at $p^{\rm (g)}(\hat{u}'), e^{\rm (g)}(\hat{u}')$. 

The evolution of the phase factor $\xi_{mn}(p(\hat{u}),e(\hat{u}))$ also changes when the secondary spin is included. The linear in spin part of the phase factor
\begin{equation}
    \delta \xi_{mn} = \eval{\pdv{\xi_{mn}}{\sigma}}_{\sigma=0} + \pdv{\xi_{mn}}{p} \delta p + \pdv{\xi_{mn}}{e} \delta e \, ,
\end{equation}
evaluated at $p^{\rm (g)}(\hat{u})$, $e^{\rm (g)}(\hat{u})$ contributes to the phase as $\sigma\, \delta \xi_{mn} \ll 1$. This contribution is of the same order as the second post-adiabatic term and can be neglected in the framework of a first order post-adiabatic analysis. Note, however, when the inspiral approaches the separatrix, our approximation fails because $\delta p$ and $\delta e$ diverge (see Sec.~\ref{sec:results}) and a different scheme must be employed.

\section{Numerical implementation and results}
\label{sec:implementationResults}

In this section we discuss how we implemented the results from the previous sections in order to calculate an inspiral of a spinning particle into a Kerr black hole in the linearized in spin approximation. Moreover, we present the phase shifts $\sigma \delta \Phi_i$ between the phase of an inspiral with a spinning secondary and inspiral with a non-spinning secondary. All the calculations were done in \emph{Mathematica} and we have used the \emph{Black Hole Perturbation Toolkit} (BHPT) \citep{BHPToolkit}.

\subsection{Implementation}
\label{sec:implementation}

Let us now discuss our approach to the numerical calculations of the adiabatic inspirals and of the phase shift in steps.
\begin{enumerate}
    \item For given $p$ and $e$, we calculate conservative trajectories, i.e. we find $\hat{E}$, $\hat{J}_z$, $\hat{\Omega}_i$, $\check{\hat{t}}(\chi), \check{\hat{r}}(\chi), \check{\phi}(\chi)$;
    \item we find the linear in $\sigma$ parts of the trajectory, i.e. $\delta p$, $\delta e$, $\delta\hat{E}$, $\delta\hat{J}_z$, $\delta \hat{t}(\chi), \delta \hat{r}(\chi), \delta \phi(\chi)$;
    \item we compute the partial amplitudes $\hat{C}^{{\rm (g)}\pm}_{lmn}$ and $\delta \hat{C}^\pm_{lmn}$ over a range of $l$, $m$ and $n$;
    \item we repeat the steps 1.-3. for many points in the $p-e$ plane and then we interpolate the total energy and angular momentum fluxes;
    \item we calculate the evolution of $p^{\rm (g)}(\hat{t})$, $e^{\rm (g)}(\hat{t})$, $\delta p(\hat{t})$, $\delta e(\hat{t})$ for given initial parameters using the interpolated fluxes;
    \item using $p^{\rm (g)}(\hat{t})$, $e^{\rm (g)}(\hat{t})$, $\delta p(\hat{t})$, $\delta e(\hat{t})$ we find the linear parts of the phases $\delta \Phi_i$.
\end{enumerate}
The above steps are described in detail in the following sections.

\subsubsection{Trajectories}

Before we calculate the amplitudes $\hat{C}^\pm_{lmn}$, we have to precompute the orbital quantities. For given $\ha$, $p$ and $e$ we calculate the geodesic quantities $\hat{\Omega}_i$, $\hat{E}$, $\hat{J}_z$, $\check{\hat{t}}(\chi),\check{\hat{r}}(\chi),\check{\phi}(\chi)$ and the linear corrections due to the secondary spin with respect to this geodesic for the same frequencies. In particular, we obtain $\delta p$ and $\delta e$ from Eqs.~\eqref{eq:deltape}, $\delta \hat{E}$ and $\delta \hat{J}_z$ we get from Eqs.~\eqref{eq:deltaE_deltaJz} and, finally, $\delta \hat{t}(\chi), \delta \hat{r}(\chi), \delta \hat{\phi}(\chi)$ are calculated from Eqs.~\eqref{eq:delta_t_phi}. Moreover, the geodesic quantities $\check{t}(\chi)$ and $\check{\phi}(\chi)$ are calculated through the BHPT, which uses the discrete cosine transform (DCT) \citep{Hopper:2015}. This method numerically transforms the integrand in Eqs.~\eqref{eq:tchi_phichi} into a series of cosines which is trivial to integrate. Actually, the linear in spin part of the trajectory, i.e. $\delta t(\chi)$ and $\delta \phi(\chi)$, is derived by employing DCT  on 50 points obtained from Eqs.~\eqref{eq:delta_t_phi}. With this number of points the error is less than $10^{-6}$ for all the calculated orbital configurations, however, note that this error is much lower for orbits far from the separatrix and for orbits with lower eccentricity.

\subsubsection{Gravitational-wave fluxes}

The obtained orbital parameters can now be used for the calculation of the partial amplitudes. The description of how to calculate the non-linearized in spin amplitudes $\check{\hat{C}}^\pm_{lmn}$ can be found in \citep{Skoupy:2021b}. In this work, we discuss the procedure allowing us to calculate the geodesic partial amplitude $\check{\hat{C}}^{\rm (g)\pm}_{lmn}$ from Eq.~\eqref{eq:Cpm_lmn_final} for $\sigma=0$ and the linear in spin part $\eval{\delta \hat{C}^\pm_{lmn}}_{\Omega_i}$ according to Eq.~\eqref{eq:deltaCpm_lmn}. In particular, the integral in \eqref{eq:deltaCpm_lmn} is evaluated using the midpoint rule, which should have exponential convergence \citep{Hopper:2015}; while for the calculation of the homogeneous solutions $R^\pm_{lm\omega}$ and $S^{a\omega}_{lm}$ the BHPT has been employed. More details about the calculation of the partial amplitudes and tests of their validity can be found in Appendix \ref{app:linearized_amplitudes}.

To obtain an adequately accurate energy or angular momentum flux, we need to calculate the amplitudes $ \mathcal{F}_{l,m,n}$ for a range of $l$, $m$ and $n$ values. Thanks to the symmetry
\begin{align}
    \mathcal{F}_{l,m,n} &= \mathcal{F}_{l,-m,-n} \, ,\\
    \hat{\omega}_{m,n} &= -\hat{\omega}_{-m,-n} \, ,
\end{align}
we decided to calculate only the modes with $\hat{\omega}_{mn}>0$ and the total sum $\mathcal{F}$ can be found as double of the sum of calculated modes\footnote{All formulas in this subsection are valid both for the fluxes $\mathcal{F}$ and their linear parts $\delta\mathcal{F}$. We demonstrate the formulas with $\mathcal{F}$ for brevity.}. The structure of the summation is
\begin{subequations}
\begin{align}
    \mathcal{F} &= 2 \sum_{m=m_{\rm min}}^{m_{\rm max}} \mathcal{F}_m \, , \\
    \mathcal{F}_m &= \sum_{l=l_{\rm min}}^{l_{\rm max}} \mathcal{F}_{lm} \, , \\
    \mathcal{F}_{lm} &= \sum_{n=n_{\rm min}}^{n_{\rm max}} \mathcal{F}_{lmn} \, ,
\end{align}
\end{subequations}
where $m_{\rm min} = -5$, $l_{\rm min} = \max\left\{2,\abs{m}\right\}$ and $m_{\rm max}$, $l_{\rm max}$, $n_{\rm min}$, $n_{\rm max}$ are chosen dynamically according to a given accuracy $\epsilon$, i.e. the maximal allowed error. This error for the geodesic fluxes should be lower than the mass ratio, otherwise it will be larger than the contribution from the post-adiabatic terms, notably the secondary spin. In our calculations we set the accuracy of the geodesic fluxes to $\epsilon=10^{-6}$ and the accuracy of the linear corrections to the fluxes to $\epsilon = 10^{-3}$.

Our first step in our computation scheme is to calculate the modes with $m=l=2$, $\lceil -m \hat{\Omega}_\phi/\hat{\Omega}_r \rceil \leq n \leq 20$, where the lower bound corresponds to the mode with minimal $n$, for which $\hat{\omega}_{mn}>0$. In all the cases we treated, the mode with maximal flux $\max_{l,m,n}\mathcal{F}_{lmn}$ lays in this range. Then we continue the summation in $n$ until the stopping condition for $n_{\rm max}$ is reached. This stopping condition is, that the magnitude of three successive modes drops below $(\epsilon/10)\max\mathcal{F}_{lmn}$. This condition must be satisfied for three consecutive modes, because the modes are not monotonic in $n$, as has been reported already in other papers \citep{Drasco:2005kz,Hughes:2021}.

At this point we have obtained the dominant $\mathcal{F}_{l=2,m=2}$ mode. Similarly we calculate the other $\mathcal{F}_{l,m=2}$ modes until the stopping condition for $l_{\rm max}$, i.e. $\mathcal{F}_{l_{\rm max} m}<\epsilon\mathcal{F}_{2,2}$, is satisfied. The magnitude of $\mathcal{F}_{lm}$ drops quickly with $l$ and usually for given $m$ no more than 4 $l$-modes are needed. In this way we obtain the dominant $\mathcal{F}_{m=2}$ mode. After that we calculate other $m$-modes. For high $m$, modes with low $n$ can be neglected. Therefore, we start the sum over $n$ at $n_0 = \lfloor 10 m e^2 \rfloor$, which is close to the maximal value of $\mathcal{F}_{lmn}$ for given $l$ and $m$ as we found empirically. Then we increase $n$ until the stopping condition for $n_{\rm max}$ is satisfied. Finally, we decrease $n$ until the condition for $n_{\rm min}$ is satisfied or until we reach $n=\lceil - m \hat{\Omega}_\phi/\hat{\Omega}_r \rceil$. 

The above procedure is repeated for other values of $m$. The stopping condition for $m_{\rm max}$ is
\begin{equation} \label{eq:condition_m}
    \frac{\mathcal{F}_{m_{\rm max}}}{1-\mathcal{F}_{m_{\rm max}}/\mathcal{F}_{m_{\rm max}-1}} < \frac{\epsilon}{2} \sum_{m=m_{\rm min}}^{m_{\rm max}} \mathcal{F}_m \, .
\end{equation}
If we assume that for high $m$ the modes $\mathcal{F}_m$ decrease exponentially, the lhs of Eq. \eqref{eq:condition_m} corresponds to the terms neglected by the truncation of the sum over $m$ at $m_{\rm max}$. For orbits with low $p$ around a Kerr black hole with $\hat{a}=0.9$ the number of $m$-modes required for an accuracy $\epsilon=10^{-6}$ is very high, so we truncate the sum at $m_{\rm max}=25$ consciously knowing that we lose in accuracy.

The amplitudes were calculated in \emph{Mathematica} using extended precision. For lower $\ha$, $l$ and $m$ the input parameters are given to 48 places. However, for modes with higher $\hat{\omega}$ and $\ha$ the calculation returns wrong result due to the loss of precision during the calculation of $R^\pm_{lmn}$. Therefore, we check if the result lays orders of magnitudes away from the Newtonian amplitudes for circular orbits in Eq.~(B3) in \citep{Nagar:2019} and when it does, we repeat the calculation with higher precision. The maximal precision is 112 places for higher $\ha$, $l$, $m$ and $n$ and lower $p$.

The calculation of individual modes with low eccentricity and $n$ takes around one second, but for high eccentricities and $n$ the computation time can be up to tens of seconds. All the modes in one grid point are calculated in around 1 hour (1 day) for lower (higher) eccentricity. The calculation of the whole grid takes hundreds of CPU hours.

\subsubsection{Interpolation in the $p-e$ plane}
\label{sec:interpolation}

Because of the high computational cost, instead of calculating the fluxes during the evolution of the orbital parameters, they are precalculated on a grid in the $p-e$ plane and then interpolated. The grid is chosen to reflect the behavior near the separatrix and to avoid some problematic regions. Actually, this grid is not in the $p$ and $e$ coordinates, but in a new set of variables $\mathsf{x}$, $\mathsf{y}$ which are obtained after several transformations from $p$ and $e$. 

The first transformation reads
\begin{align}
    \tilde{U} &= \sqrt{(p-\hr_{\rm ISCO})^2 - (p_{\rm s}(e) - \hr_{\rm ISCO})^2} \label{eq:p2u} \, , \\
    V &= e^2 \, ,
\end{align}
where $p_{\rm s}(e)$ is the location of the separatrix. The purpose of this transformation is to make the quantities and their derivatives finite for circular orbits, i.e. for $e=0$. Namely, since the fluxes depend only on even powers of $e$, their derivative with respect to $e$ vanishes for $e=0$. The inverse relation of Eq.~\eqref{eq:p2u} reads
\begin{equation}
    p = \hr_{ISCO} + \sqrt{\tilde{U}^2 + (p_{\rm s}(\sqrt{V})-\hr_{ISCO})^2} \; .
\end{equation}

Next, we transform from $\tilde{U}$ to
\begin{equation}
    U = \frac{c}{\log(1+c/\tilde{U})}
\end{equation}
to regularize the quantities near the separatrix. $c$ is a parameter controlling the grid density near the separatrix. For higher $c$ the grid points are more dense near the separatrix while for $c\rightarrow 0$ it holds $U\rightarrow \tilde{U}$. We have chosen the value $c=25$ in our calculations. The asymptotic behavior of these transformations is
\begin{enumerate}
    \item $U \rightarrow p$, when $p \rightarrow \infty$, and
    \item  $U \rightarrow -1/\log(p-p_{\rm s})$, when $p\rightarrow p_{\rm s}$, 
\end{enumerate}
which is proportional to the behavior of the radial frequency $\hat{\Omega}_r$ near the separatrix \citep{Glampedakis:2002,Warburton:2013}.

We made one additional transformation to avoid two areas with high eccentricity: a) an area with high $p$, for which the total time of the inspiral is very long, and b) an area close to the separatrix, for which the inspiral must start with very high eccentricity.  This transformation to $\mathsf{x} \in (0,1)$, $\mathsf{y} \in (0,1)$ is given by
\begin{align}
    U &= (U_{11}-U_{10}+U_{00}-U_{01}) \mathsf{x} \mathsf{y} + (U_{10}-U_{00}) \mathsf{x} \nonumber \\ &\phantom{=} + (U_{01} - U_{00}) \mathsf{y} + U_{00} \, , \\
    V &= (V_{11}-V_{01}) \mathsf{x} \mathsf{y} + (V_{01}-V_{00}) \mathsf{y}  \, ,
\end{align}
where the parameters $U_{\mathsf{xy}}$, $V_{\mathsf{xy}}$ are chosen according to the boundaries described in the following paragraph.

\begin{figure}[htb!]
    \centering
    \includegraphics[width=0.46\textwidth]{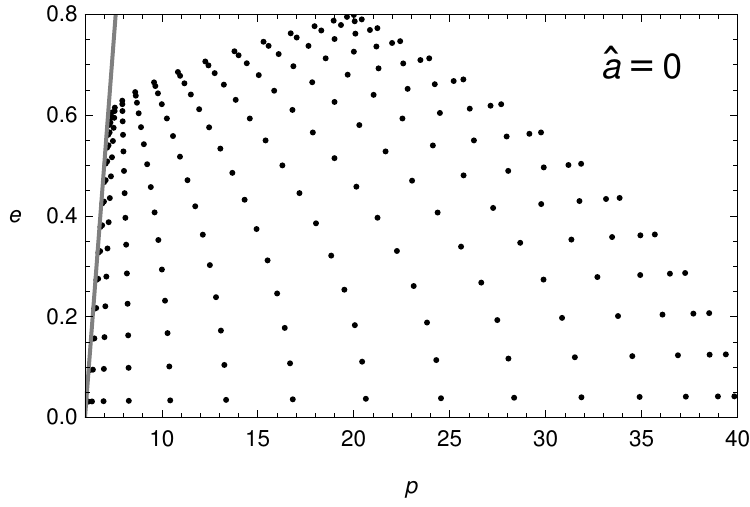}
    \includegraphics[width=0.46\textwidth]{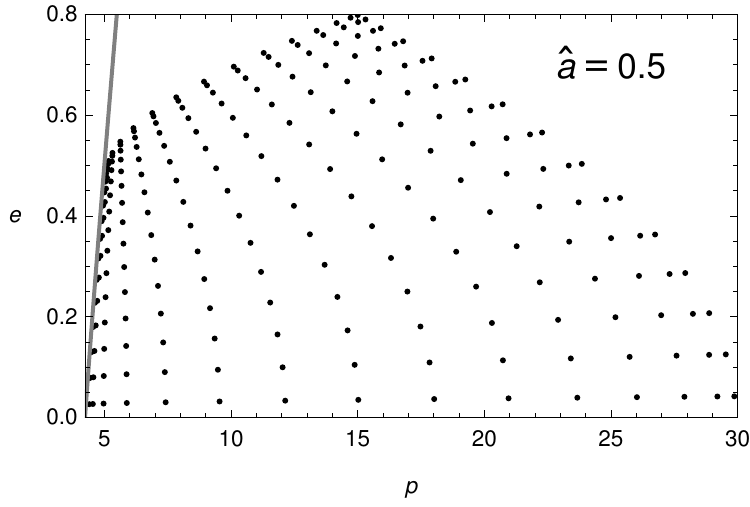}
    \includegraphics[width=0.46\textwidth]{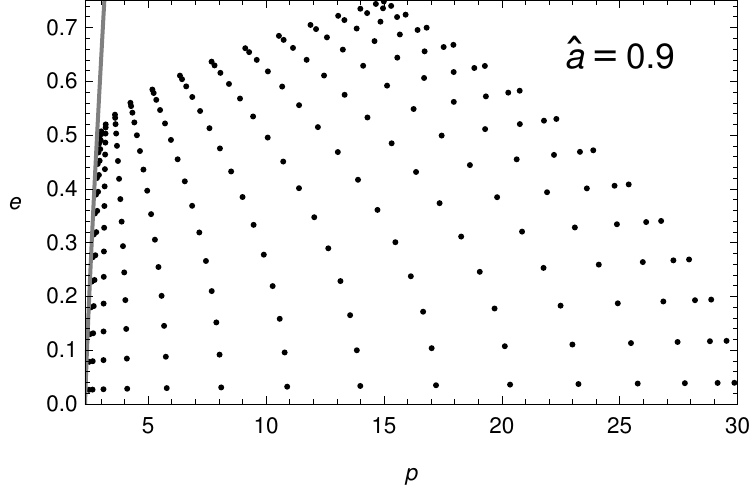}
    \caption{Grids for the interpolation in $p-e$ plane. The grid points are at Chebyshev nodes in $\mathsf{x}-\mathsf{y}$ plane}
    \label{fig:grids}
\end{figure}

The GW fluxes were calculated on a grid in Chebyshev nodes in the $\mathsf{x}$, $\mathsf{y}$ coordinates. We used 15 grid points in both directions. The boundaries were chosen for each value of $\ha$ separately. In all $\ha$ cases the coordinates of the lower left corners are $(p,e) = (\hr_{\rm ISCO}(\ha)+0.15,0)$. For $\ha=0$ the upper left corner is at $(p_{\rm s}(0.6)+0.1,0.6)$ and at $(p_{\rm s}(0.5)+0.1,0.5)$ for $\ha=0.5,0.9$. The lower right corner is located at $(40,0)$ or $(30,0)$ for $\ha=0$ or $\ha=0.5,0.9$ respectively. The coordinates of the upper right corner are $(20,0.8)$ for $\ha=0$, $(15,0.8)$ for $\ha=0.5$ and $(15,0.75)$ for $\ha=0.9$. These grids are depicted in Fig.~\ref{fig:grids}.

On the grid we interpolated the total energy and angular momentum fluxes $\mathcal{F}^{E {\rm (g)}}$, $\mathcal{F}^{J_z {\rm (g)}}$ with their linear in spin counterparts $\delta \mathcal{F}^{E}$, $\delta\mathcal{F}^{J_z}$, the time derivatives of the orbital parameters $\dot{p}^{\rm (g)}$, $\dot{e}^{\rm (g)}$ and the derivatives of $\dot{p}$ and $\dot{e}$ with respect to $\sigma$, $p$ and $e$ for the calculation of $\delta\dot{p}$ and $\delta\dot{e}$ using Eq.~\eqref{eq:ddeltapedt}. Each function was divided by the following normalization factors to regularize the behavior near the separatrix, for high $p$ and for low $e$:
\begin{subequations}
\begin{align}
    N_{\mathcal{F}^E} &= \frac{32}{5} p^{-5} \qty(1-e^2)^{3/2} \qty(1+\frac{73}{24} e^2 + \frac{37}{96} e^4) \, , \\
    N_{\mathcal{F}^{J_z}} &= \frac{32}{5} p^{-7/2} \qty(1-e^2)^{3/2} \qty(1+\frac{7}{8}e^2) \, , \\
    N_{\delta\mathcal{F}^E} &= - \frac{25}{4} p^{-3/2} N_{\mathcal{F}^E} \frac{p^2}{\tilde{U}^2} \, ,\\
    N_{\delta\mathcal{F}^{J_z}} &= - \frac{25}{4} p^{-3/2} N_{\mathcal{F}^{J_z}} \frac{p^2}{\tilde{U}^2} \, , \\
    N_{\dot{p}^{\rm (g)}} &= \frac{8}{5} p^{-3} \qty(1-e^2)^{3/2} \qty( 8 + 7e^2 ) \frac{p^2}{\tilde{U}^2} \, , \\
    N_{\dot{e}^{\rm (g)}} &= \frac{1}{15} e p^{-4} \qty(1-e^2)^{3/2} \qty( 304 + 121e^2 ) \frac{p^2}{\tilde{U}^2} \, , \\
    N_{\partial_{\sigma} \dot{p}} &= \frac{1}{\tilde{U}^{4}} \qty(1-e^2)^{3/2} \, , \\
    N_{\partial_{\sigma} \dot{e}} &= e \frac{1}{p \tilde{U}^{4}} \qty(1-e^2)^{3/2} \, ,
\end{align}
\begin{align}
    N_{\partial_p \dot{p}} &= \frac{p}{\tilde{U}^{4}} \qty(1-e^2)^{3/2} \, , \\
    N_{\partial_e \dot{p}} &= e \frac{p^2}{\tilde{U}^{4}} \, ,\\
    N_{\partial_p \dot{e}} &= e \frac{1}{\tilde{U}^{4}} \qty(1-e^2)^{3/2} \, , \\
    N_{\partial_e \dot{e}} &= \frac{p}{\tilde{U}^{4}} \, .
\end{align}
\end{subequations}
The behavior of $N_{\mathcal{F}^E}$ and $N_{\mathcal{F}^{J_z}}$ comes from \citep{Peters:1963}, where they derived the fluxes from a keplerian orbit, which represents the large $p$ limit. On the other hand, the behavior of $N_{\delta\mathcal{F}^E}$ and $N_{\delta\mathcal{F}^{J_z}}$ for large $p$ is derived from the post-Newtonian GW fluxes of spinning particles on circular equatorial orbits \citep{Tanaka:1996ht}. The accuracy of the interpolation is discussed in Appendix \ref{app:accuracyInt}.

\subsubsection{Evolution of the orbital parameters}

By using the interpolated functions obtained in the previous section multiplied by the normalization factors allows the calculation of the evolution of the geodesic orbital parameters $p^{\rm (g)}(\hat{t})$, $e^{\rm (g)}(\hat{t})$ and the respective corrections $\delta p(\hat{t})$, $\delta e(\hat{t})$. For given initial parameters $p_0^{\rm (g)}$ and $e_0^{\rm (g)}$ we numerically solved the equations~\eqref{eq:dp0e0dt} in \emph{Mathematica} using the $7/8$th order Runge-Kutta method with adaptive step-size. The calculation were terminated when the orbital parameters reached the boundary at ${\mathsf x}=0$.

These results were then used to evolve Eqs.~\eqref{eq:ddeltapedt} for given initial conditions $\delta p_0^{\rm (g)}$ and $\delta e_0^{\rm (g)}$. These initial conditions specify the trajectory of a spinning particle, which is then compared with the geodesic starting at $p_0^{\rm (g)}$ and $e_0^{\rm (g)}$. The case $\delta p_0 = 0 = \delta e_0$ corresponds to a trajectory of a spinning particle compared with a geodesic which starts at the same $p_0^{\rm (g)}$ and $e_0^{\rm (g)}$. 

However, $\delta p_0$ and $\delta e_0$ can be chosen such that we compare a trajectory of a spinning particle with a geodesic with the same initial orbital frequencoes $\hat{\Omega}_r$ and $\hat{\Omega}_\phi$. In this case, we set $\delta p_0$ and $\delta e_0$ to
\begin{subequations}\label{eq:initial_Omega_i}
\begin{align}
    \delta p_0 = \delta p(p_0^{\rm (g)}, e_0^{\rm (g)}) \, , \\
    \delta e_0 = \delta e(p_0^{\rm (g)}, e_0^{\rm (g)}) \, ,
\end{align}
\end{subequations}
where the functions $\delta p$ and $\delta e$ have been defined in \eqref{eq:deltape}.

We have also calculated the case, where the trajectory of a spinning particle is compared with a geodesic with the same initial eccentricity $e$ and azimuthal frequency $\hat{\Omega}_\phi$. This choice was used in previous works \citep{Piovano:2020,Skoupy:2021a} when calculating quasicircular inspirals. In this case, we set
\begin{subequations}\label{eq:initial_Omega_phi_e}
\begin{align}
    \delta p_0 &= - \frac{\pdv{\hat{\Omega}_\phi}{\sigma}}{\pdv{\hat{\Omega}_\phi}{p}} \\
    \delta e_0 &= 0
\end{align}
\end{subequations}
evaluated at $p_0^{\rm (g)}$, $e_0^{\rm (g)}$ and $\sigma=0$.

\subsubsection{Evolution of the phase shifts}

After the calculation of the orbital parameters we calculated the linear parts of the phases $\delta \Phi_i$ using Eq.~\eqref{eq:deltaPhi} with the default solver \texttt{NDSolve} in \emph{Mathematica}. The results were compared with non-linearized inspiral to verify them. Details are given in Appendix \ref{app:accuracyPhases}.

\subsection{Results}
\label{sec:results}

\subsubsection{Matched eccentricity and azimuthal frequency}
\label{sec:mathedOmega_phi_e}

\begin{figure}
    \centering
    \includegraphics[width=0.45\textwidth]{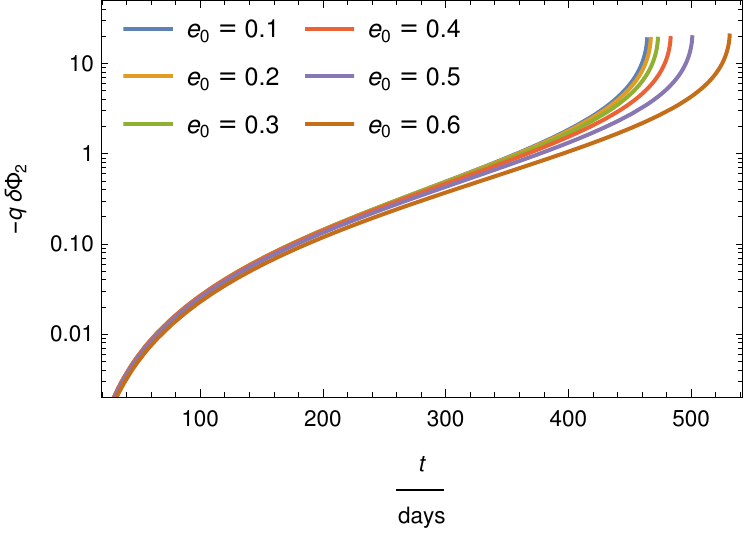}
    \caption{The phase shift $q \delta\Phi_m = q m\delta\Phi_\phi$ of the dominant $m=2$ mode for properly matched initial azimuthal frequency $\Omega_\phi$ and eccentricity $e$. The inspirals of a spinning particle with $\mu=30M_\odot$ into a Kerr black hole with $M = 10^6 M_\odot$, $\ha=0.9$ start from $p^{\rm (g)}_0=10.1$.}
    \label{fig:deltaPhi_phi_circ}
\end{figure}

When the phase shift $\delta \Phi_\phi$ is calculated for circular orbits, the phase from an inspiral with a non-spinning secondary is compared with an inspiral with a spinning secondary which has the same initial azimuthal frequency $\Omega_\phi$ and initial eccentricity $e=0$ as the inspiral with non-spinning secondary. Obviously the radial frequency $\Omega_r$ is not relevant for circular orbits, in fact, the partial amplitudes $C^0_{lmn}$ vanish for $n \neq 0$ and only the modes with frequency $m\Omega_\phi$ remain. However, we can extend this approach to the calculation of the phase shift from eccentric inspirals by choosing properly the initial conditions as given in Eq.~\eqref{eq:initial_Omega_phi_e}. The corresponding numerical examples are given in Fig.~\ref{fig:deltaPhi_phi_circ}, which shows the phase shift $\delta\Phi_{2,0}$ for the dominant $m=2$ mode. Fig.~\ref{fig:deltaPhi_phi_circ} is consistent with Fig.~2 from \citep{Skoupy:2021a} and Fig.~3 from \citep{Piovano:2020}. Note that since we examine the phase at constant distance from the central black hole, i.e. at constant $\hr$, we can use $t$ as the time variable instead of $u$. 

\begin{figure}
    \centering
    \includegraphics[width=0.45\textwidth]{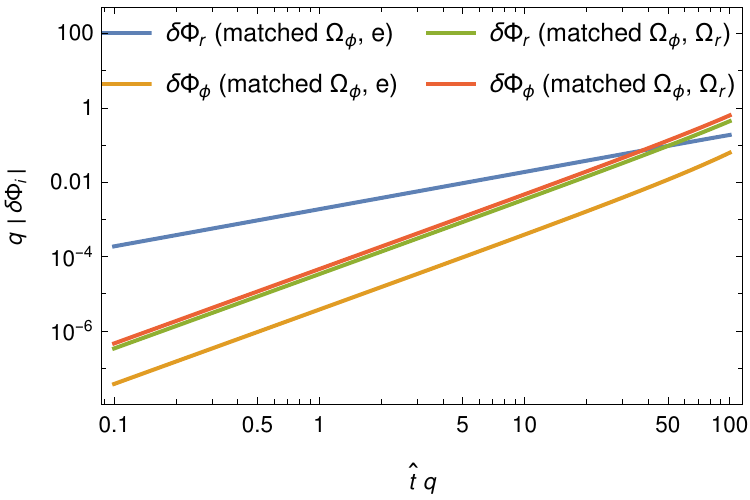}
    \caption{Phase shifts $q \delta\Phi_i$ for different initial conditions. The inspirals are around a Kerr black hole with $\ha=0.9$ and start at $p^{\rm (g)}_0=10.1$ and $e^{\rm (g)}_0=0.2$. For matched $\Omega_\phi$ and $e$, $\delta\Phi_\phi$ grows as $\hat{t}^2$ and $\delta\Phi_r$ grows as $\hat{t}$ for low $t$, while for matched $\Omega_r$ and $\Omega_\phi$ both $\delta\Phi_r$ and $\delta\Phi_\phi$ grow as $\hat{t}^2$.}
    \label{fig:deltaPhi_i_circ}
\end{figure}

When the initial azimuthal frequency and eccentricity are properly matched, the phase shift $\delta \Phi_\phi$ grows as $\hat{t}^2$, whereas $\delta\Phi_r$ grows as $\hat{t}$ for low $\hat{t}$, as can be seen in Fig.~\ref{fig:deltaPhi_i_circ}. The reason for this behavior is that the initial value for
\begin{equation}
    \delta \hat{\Omega}_i = \eval{\pdv{\hat{\Omega}_i}{\sigma}}_{\sigma=0} + \pdv{\hat{\Omega}_i^{\rm (g)}}{p} \delta p + \pdv{\hat{\Omega}_i^{\rm (g)}}{e} \delta e \, ,
\end{equation}
which appears in the integral \eqref{eq:deltaPhi}, is zero for $\delta\hat{\Omega}_\phi$, but it is non-zero for $\delta\hat{\Omega}_r$. Thus, the phase shift $\delta\Phi_r$ grows linearly in $\hat{t}$ after the integration for low $\hat{t}$.

\subsubsection{Matched frequencies}

\begin{figure}
    \centering
    \includegraphics[width=0.45\textwidth]{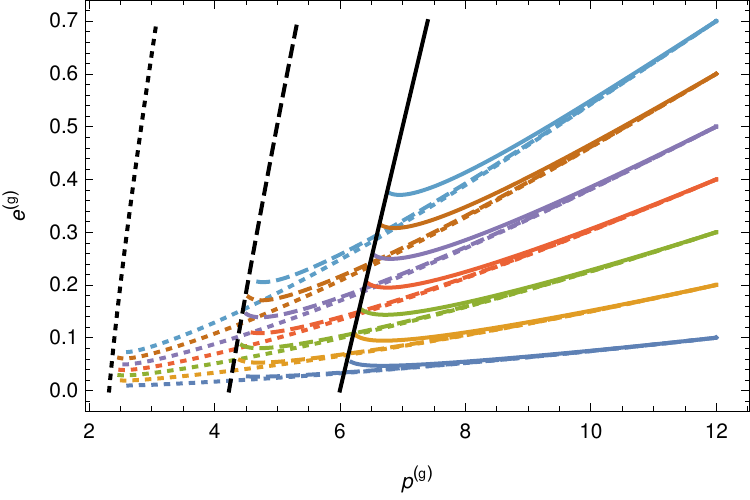}
    \caption{Adiabatic evolution of $p^{\rm (g)}$ and $e^{\rm (g)}$ for $\ha=0$ (solid), $\ha=0.5$ (dashed) and $\ha=0.9$ (dotted), while the respective black lines denote the separatrices, where the evolution ends.}
    \label{fig:adiabatic}
\end{figure}

\begin{figure}
    \centering
    \includegraphics[width=0.45\textwidth]{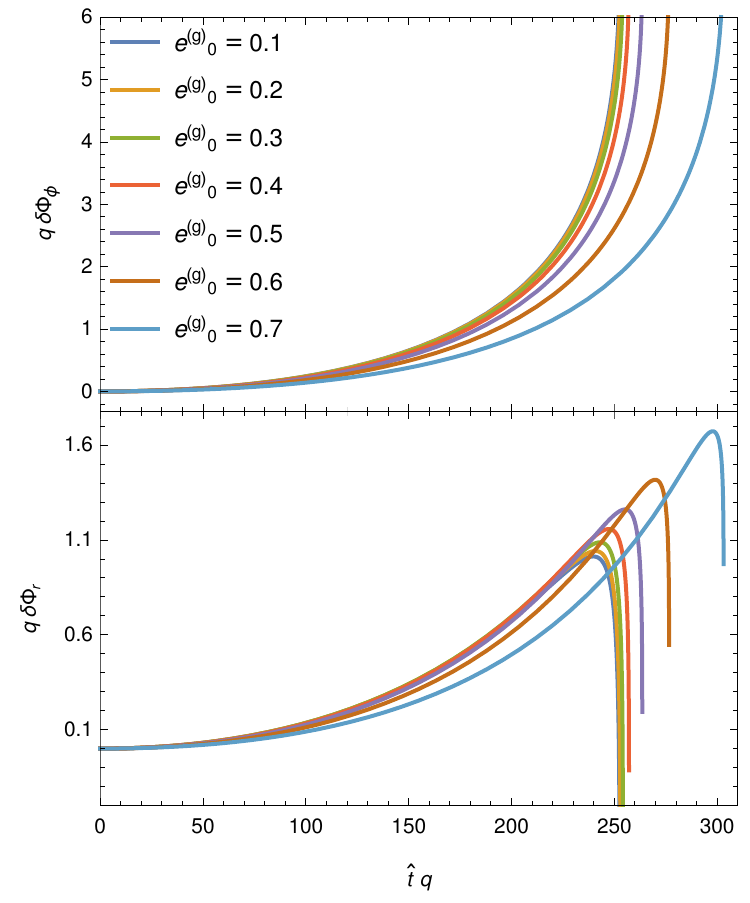}
    \caption{The azimuthal (top) and the radial (bottom) phase shift for orbits around a Schwarzschild black hole with initial semi-latus rectum $p^{\rm (g)}_0=12$ and different initial eccentricities. This plot shows the phase shift when the particle has spin $\sigma = q$, i.e. the secondary corresponds to an extremal Kerr black hole.}
    \label{fig:deltaPhi_i_a0.0_p12.0}
\end{figure}

\begin{figure}
    \centering
    \includegraphics[width=0.45\textwidth]{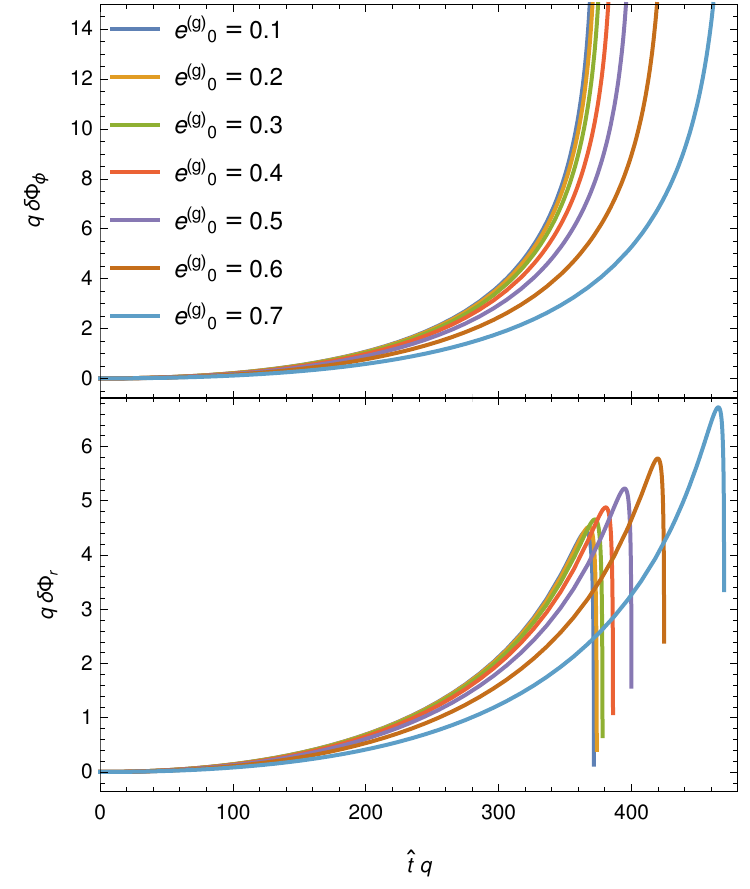}
    \caption{The same as Fig.~\ref{fig:deltaPhi_i_a0.0_p12.0}, but for a Kerr black hole with $\ha=0.5$}
    \label{fig:deltaPhi_i_a0.5_p12.0}
\end{figure}

\begin{figure}
    \centering
    \includegraphics[width=0.45\textwidth]{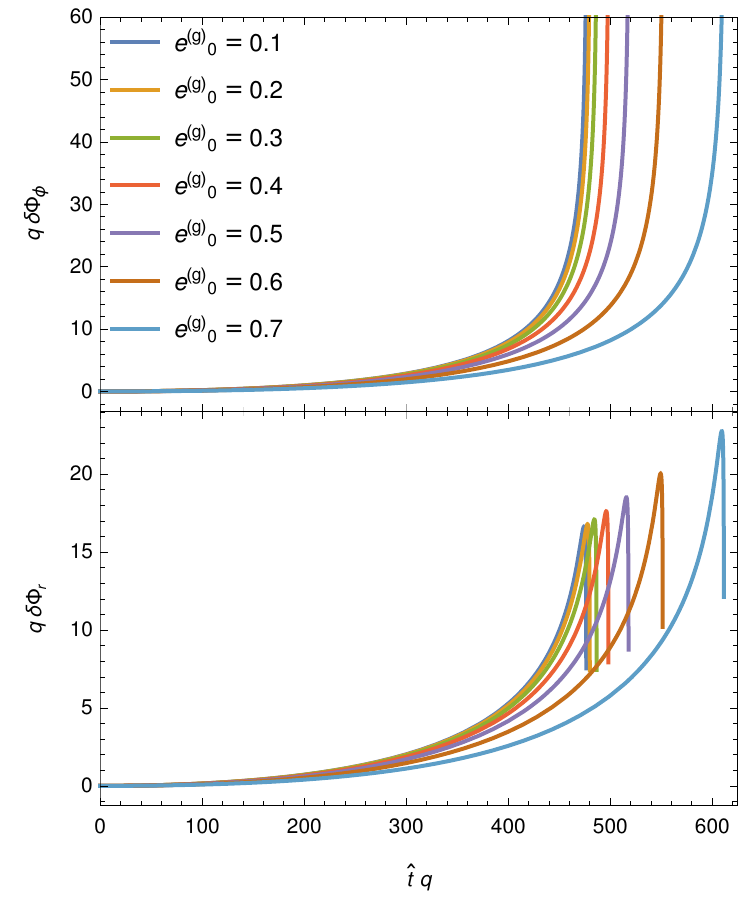}
    \caption{The same as Fig.~\ref{fig:deltaPhi_i_a0.0_p12.0}, but for a Kerr black hole with $\ha=0.9$}
    \label{fig:deltaPhi_i_a0.9_p12.0}
\end{figure}

Since for eccentric orbits both frequencies are observable, we prefer to match the initial frequencies according to Eqs.~\eqref{eq:initial_Omega_i} than as discussed in Sec.~\ref{sec:mathedOmega_phi_e}. For this initial setting both $\delta \Phi_\phi$ and $\delta \Phi_r$ grow as $\hat{t}^2$ for low $t$, as can be seen in Fig.~\ref{fig:deltaPhi_i_circ}. In the numerical example given in Fig.~\ref{fig:adiabatic} we have calculated the inspiral providing the evolution of $p^{\rm (g)}(\hat{t})$ and $e^{\rm (g)}(\hat{t})$ for initial semi-latus rectum $p^{\rm (g)}_0=12$ and different initial eccentricities. The respective phase shifts for $\ha=0$, $\ha=0.5$ and $\ha=0.9$ are shown in Figs.~\ref{fig:deltaPhi_i_a0.0_p12.0}-\ref{fig:deltaPhi_i_a0.9_p12.0}. The linear in spin part of the azimuthal phase $\delta\Phi_\phi$ is increasing and is positive as opposed to the case with matched initial $\Omega_\phi$ and $e$ in Sec.~\ref{sec:mathedOmega_phi_e}, where it is negative (see Fig.~\ref{fig:deltaPhi_phi_circ}). The linear part of the radial phase $\delta\Phi_r$ is increasing and positive for the majority of the inspiral, however, right before the trajectory reaches the separatrix, $\delta\Phi_r$ starts to decrease. Both $\delta\Phi_\phi$ and $\delta\Phi_r$ diverge when the trajectory is approaching the separatrix, because both the linearization in spin and the two-scale approximation break at the separatrix.

\begin{figure}
    \centering
    \includegraphics[width=0.45\textwidth]{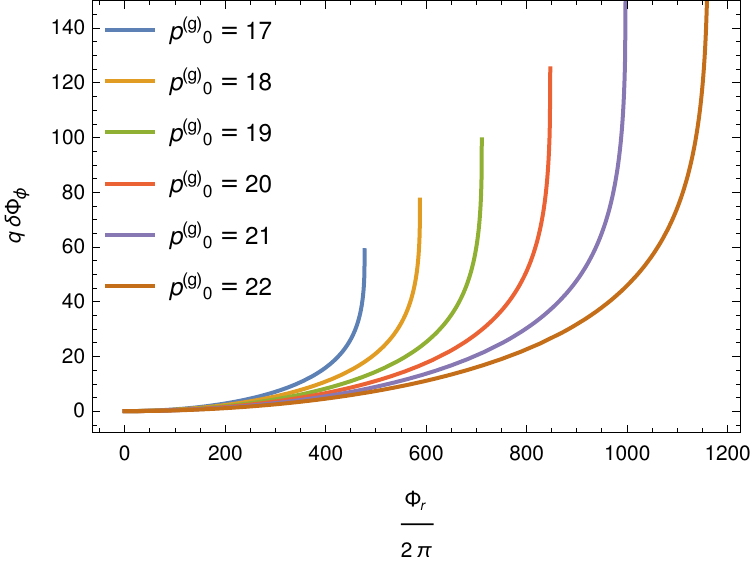}
    \caption{The phase shift for inspirals around a Schwarzschild black hole with initial eccentricity $e^{\rm (g)}_0=0.75$ and $\sigma=q$. The horizontal axis corresponds to the number of passages through the pericenter.}
    \label{fig:deltaPhi_phi_e0.75}
\end{figure}

In \citep{Warburton:2017}, where they compared eccentric equatorial inspirals of spinning particles into a Schwarzschild black hole using osculating geodesics method, they found initial parameters, for which the difference $\Delta\phi=\phi_{\sigma=q}-\phi_{\sigma=0}$ between the azimuthal coordinates $\phi_{\sigma=q}$ of a spinning body and $\phi_{\sigma=0}$ of a nonspinning body changes its sign during the inspiral (Fig.~2 in \citep{Warburton:2017}). However, that work included only the MPD force into the equations of motion and did not take into account the correction to the self-force caused by the body's spin. We have calculated the phase shift $q \delta\Phi_\phi$, which should correspond to $\Delta\phi$ when the particle passes the pericentre, for the same initial parameters with \citep{Warburton:2017} and found no change in the sign of $\Delta\phi$ (see Fig.~\ref{fig:deltaPhi_phi_e0.75}). However, note that we have not included the conservative and oscillating dissipative parts of the self-force and, thus, these results are not directly comparable. Also, the accumulated phase shift is higher in our Fig.~\ref{fig:deltaPhi_phi_e0.75}, where the secondary's spin contribution is incorporated to the fluxes, than in Fig.~2 of \citep{Warburton:2017}, where this contribution has not been taken into account.

\begin{figure}
    \centering
    \includegraphics[width=0.45\textwidth]{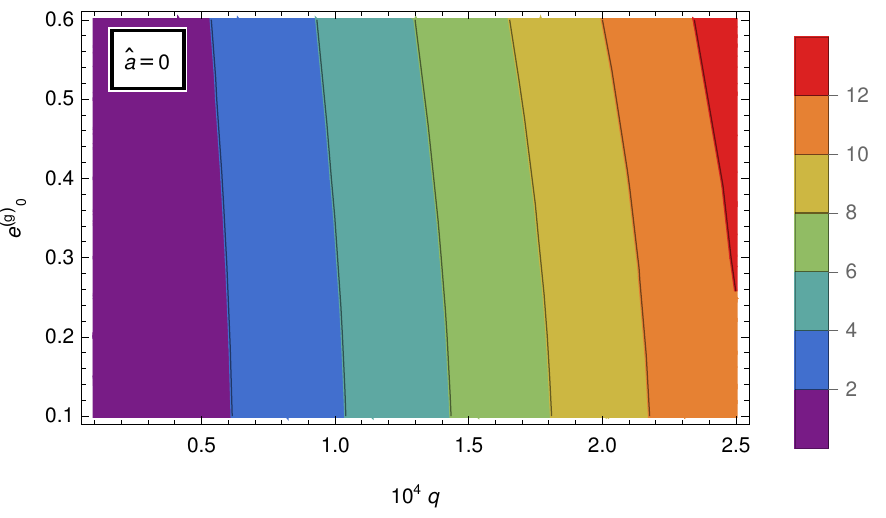}
    \includegraphics[width=0.45\textwidth]{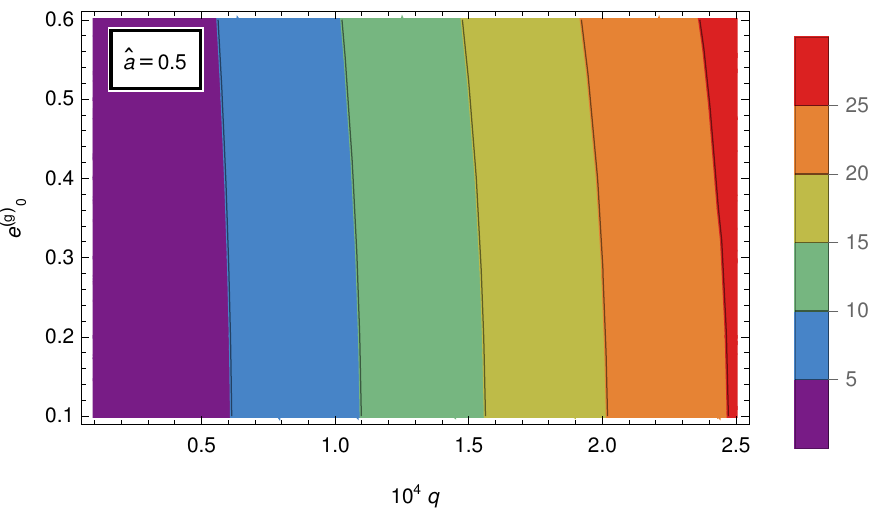}
    \includegraphics[width=0.45\textwidth]{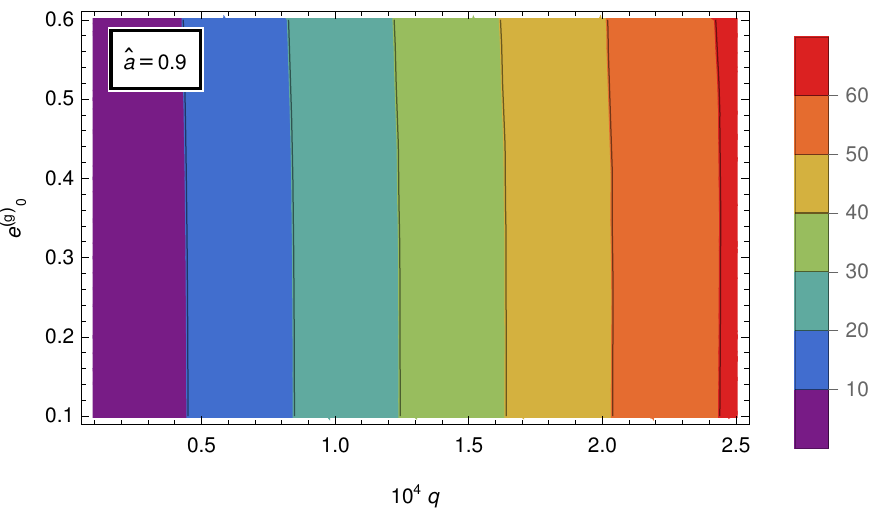}
    \caption{The maximal radial phase shift $\max{q \delta\Phi_r}$ for different initial eccentricities, mass ratios and the Kerr parameters. The mass of the central black hole is $M=10^6 M_\odot$ and the duration of the inspirals is $1$ year. This phase shift corresponds to a particle with spin $\sigma=q$.}
    \label{fig:maxdeltaPhi_r}
\end{figure}

To systematically probe the parameter space, we have calculated the inspirals for various initial parameters and for each inspiral we have found the maximum of the radial phase shift $\max q \delta\Phi_r$\footnote{In Appendix \ref{app:accuracyPhases} we verify that the accuracy of the phase shift is high and the approximations are valid at this point.}. Then we have plotted this maximum against the initial eccentricity $e^{\rm (g)}_0$ and the mass ratio $q$, assuming that the duration of the inspiral is $1$ year while the mass of the central black hole is $M=10^6 M_\odot$. At each point in the $q-e^{\rm (g)}_0$ plane the inspirals start at different initial semi-latus rectum $p^{\rm (g)}_0$. The calculation was repeated for $\ha=0,0.5,0.9$ and the results are shown in Fig.~\ref{fig:maxdeltaPhi_r}. We can see that for higher mass ratio the maximal phase shift is higher which corresponds to higher $p^{\rm (g)}_0$. For higher $\ha$ the maximal phase shift is almost independent of the initial eccentricity, but to find the degeneracies in the parameter space and to assess the detectability of the initial eccentricity or the secondary spin, proper analysis must be done, which is out of the scope of the present technical work.

\section{Conclusions} \label{sec:Concl}

We studied the influence of the spin $\sigma$ of secondary body on the phase of a GW from an EMRI moving on the equatorial plane of a Kerr black hole. Thanks to the fact, that the spin $\sigma$ is of the same order as the mass ratio $q$, we worked in the linear order in $\sigma$ neglecting higher order terms. We emphasize that our results are not sufficient for the generation of the waveform templates for the detection, since they must be accurately and rapidly generated in the whole parameter space. The purpose of this work is to provide the technical background needed to calculate the secondary's spin contributions to the waveform.

The first step to achieve our goal was to derive the linear in $\sigma$ parts of the orbital parameters $p$, $e$, constants of motion $E$, $J_z$ and the coordinate functions $t(\chi)$, $r(\chi)$, $\phi(\chi)$ in the Darwin parametrizaion. The linearization was done with respect to a reference geodesic with the same frequencies $\Omega_r$ and $\Omega_\phi$. Then we used these quantities to linearize the GW fluxes to infinity and through the horizon. We provided the linear parts $\delta\mathcal{F}^{E}$, $\delta\mathcal{F}^{J_z}$ of the total energy and angular momentum flux using the Teukolsky formalism in the frequency domain. Again, we calculated the linear part with respect to a geodesic with the same frequencies. We also found the relation between the latter type of linearization and the linearization with respect to a geodesic with the same orbital parameters $p$ and $e$. 

The fluxes were calculated on a grid in the $p-e$ plane and interpolated, since the calculation at one point is computationally expensive. Once we have calculated the energy and momentum fluxes linearized in $\sigma$, we derived the evolution equations for the orbital parameters $p^{\rm (g)}(t)$ and $e^{\rm (g)}(t)$ for non-spinning secondary and for corrections due to the spin $\delta p(t)$ and $\delta e(t)$. After that we have evolved these quantities numerically. From the evolution of the orbital parameters and their corrections we then constructed the evolution of the phase shifts $\delta\Phi_r(t)$ and $\delta\Phi_\phi(t)$, which is the difference between the GW phase from an inspiral with spinning and non-spinnig secondary. We tested the results against non-linearized evolution obtained from the fluxes, that were derived in \citep{Skoupy:2021b}. We found that the error of the phase shifts is around $10^{-3}$.

The phase shifts were computed using two different types of initial conditions. First we set the initial conditions such that we compared inspirals with spinning and non-spinning secondary which start with the same azimuthal frequency $\Omega_\phi$ and eccentricity $e$. This was done to validate the results against quasicircular inspirals. We have found the expected behavior where the azimuthal phase shift grows as $t^2$ for low $t$ and the radial phase shift grows as $t$. After that we set the initial condition such that we compare inspirals with the same initial radial frequency $\Omega_r$ and azimuthal frequency $\Omega_\phi$. We found that the azimuthal phase shift is positive, as opposed to the previous choice of initial condition, and that the radial phase shift is positive and increasing up to a point before it reaches the separatrix, where it becomes decreasing. Both the azimuthal and radial phase shift diverge when the inspiral reaches the separatrix and, thus, a different method must be employed for the waveform generation near the plunge in the future. 

To systematically probe the parameter space and find the general behavior of the phase shifts, we calculated the maximal value of the radial phase shift for different initial eccentricities, mass ratios and Kerr parameters while fixing the masses of the bodies and the observation time. We found that the maximal radial phase shift grows with the mass ratio and the Kerr parameter and almost does not depend on the eccentricity.

In the future this work can be extended to off-equatorial orbits with precessing spin, which is significantly more complex since the equation of motion are not separable, even in the linear in spin order \citep{Witzany:2019}. We are also planing to generate the waveforms using the \emph{FastEMRIWaveforms} package \citep{Katz:2021} to find the degeneracies in the parameter space and to assess the detectability, since in \citep{Piovano:2021} it was claimed that for quasicircular orbits the secondary spin should not be detectable. 


\begin{acknowledgments}
The authors have been supported by the fellowship Lumina Quaeruntur No. LQ100032102 of the Czech Academy of Sciences. VS acknowledges support by the project "Grant schemes at CU" (reg.no. CZ.02.2.69/0.0/0.0/19\_073/0016935). We would like to thank Tom\'{a}\v{s} Ledvinka and Maarten van de Meent for useful discussions and comments. This work makes use of the Black Hole Perturbation Toolkit. Computational resources were supplied by the project "e-Infrastruktura CZ" (e-INFRA CZ LM2018140 ) supported by the Ministry of Education, Youth and Sports of the Czech Republic.
\end{acknowledgments}

\appendix

\section{List of dimensionless quantities}
\label{app:dimensionless}

In this work we define some quantities in their dimensionless form. However, since we use these quantities often in both full and dimensionless form, we present the respective relations in Table~\ref{tab:dimensionless}.
\begin{table}[ht!]
    \centering
    \caption{List of dimensionless quantities}
    \begin{tabular}{r@{$\,=\,$}l|l}
        $\sigma$ & $S/(\mu M)$ & Secondary spin \\
        $\hat{t}$ & $t/M$ & BL time \\
        $\hr$ & $r/M$ & BL radial coordinate \\
        $\hat{E}$ & $E/\mu$ & Energy \\
        $\hat{J}_z$ & $J_z/(\mu M)$ & Angular momentum \\
        $\ha$ & $a/M$ & Kerr parameter \\
        $\hat{T}_r$ & $T_r/M$ & Radial period \\
        $\hat{\Omega}_r$ & $\Omega_r M$ & Radial BL frequency \\
        $\hat{\Omega}_\phi$ & $\Omega_\phi M$ & Azimuthal BL frequency \\
        $\hat{\omega}$ & $\omega M$ & Frequency \\
        $\hat{C}^\pm_{lmn}$ & $C^\pm_{lmn}M^2/\mu$ & Partial amplitudes \\
        $\hat{A}^\pm_{lmn}$ & $A^\pm_{lmn}/M$ & Waveform amplitudes \\
        $\hat{u}$ & $u/M$ & Retarded coordinate \\
        $\hat{\Delta}$ & $\Delta/M^2$ & \\
        $\hat{\varpi}^2$ & $\varpi^2/M^2$ &
    \end{tabular}
    \label{tab:dimensionless}
\end{table}

Note that: some quantities such as $x$ or the fluxes $\mathcal{F}$ have been defined solely as dimensionless; quantities derived from others, e.g. by linearization in $\sigma$, have the same relation between their dimensionless and full form as the original quantities.

\section{Eccentric equatorial orbits of spinning particles}
\label{app:trajectory}

This Appendix briefs some formulas describing the motion of spinning particles on bound eccentric equatorial orbits around a Kerr black hole. Details regarding these formulas can be found in \citep{Saijo:1998,Skoupy:2021b}. 

Bound equatorial orbits of a spinning particle moving around a Kerr black hole can be parametrized by the eccentricity $e$ and the semi-latus rectum $p$. This parametrization is in one-to-one correspondence to the parametrization with respect to the energy $\hat{E}$ and the $z$-component of total angular momentum $\hat{J}_z$. The expressions of $\hat{E}$ and $\hat{J_z}$ as functions of $p$ and $e$ read
\begin{align} \label{eq:energy2}
    \hat{E}^2 = \frac{\kappa\rho + 2\epsilon\tilde{\sigma} - 2 \sgn{(\hat{J}_z)}\, \tilde{\sigma}\sqrt{\epsilon^2 + \kappa \zeta}}{\rho^2 + 4\eta\tilde{\sigma}} \, , \\
    \hat{J}_z = \frac{\epsilon\rho - 2\kappa\eta - \sgn{(\hat{J}_z)}\,\rho \sqrt{\epsilon^2 + \kappa \zeta}}{(\rho^2 + 4\eta\tilde{\sigma}) \hat{E}}\; , \label{eq:angmom2}
\end{align}
where the coefficients
\begin{align*}
\kappa &= d_1 h_2 - d_2 h_1 \; , \\
\epsilon &= d_1 g_2 - d_2 g_1 \; ,\\
\rho &= f_1 h_2 - f_2 h_1 \; , \\
\eta &= f_1 g_2 - f_2 g_1 \; ,\\
\tilde\sigma &= g_1 h_2 - g_2 h_1 \; , \\
\zeta &= d_1 f_2 - d_2 f_1 \;
\end{align*}
are calculated from the functions
\begin{align*}
f(\hr) =& \ha^2 (\hr+2) \hr + \hr^4 +\\  &+ \sigma \left(\frac{\ha^2 \sigma}{\hr^2}+\frac{2 \ha^2 (\ha + \sigma)}{\hr} + 6 \ha \hr-(\hr-2) \hr \sigma \right) \; , \\
g(\hr) =& 2 \ha \hr + \sigma \left(\frac{\ha \sigma}{\hr^2} + \frac{\ha (2 \ha + \sigma)}{\hr}-(\hr-3) \hr\right) \; , \\
h(\hr) =& \hat{\Delta} - \left(\ha+\frac{\sigma}{\hr}\right)^2 \; ,\\
d(\hr) =& \frac{\hat{\Delta} (\hr^3 - \sigma^2)^2}{\hr^4} \; . 
\end{align*}
at the pericenter $f_1=f(p/(1+e))$ and at the apocenter $f_2=f(p/(1-e))$ etc.

The trajectories in Darwin parametrization can then be calculated from the evolution equations \eqref{eq:EOM_tphi} with
\begin{align}
    V^t &= \ha\left( 1+\frac{3\sigma^2}{\hr \Sigma_\sigma} \right) x + \frac{\varpi^2}{\Delta} P_\sigma \; , \\
    V^\phi &= \left( 1+\frac{3\sigma^2}{\hr \Sigma_\sigma} \right) x + \frac{\ha}{\hat{\Delta}} P_\sigma \; , \\
    P_\sigma &= \Sigma_\sigma \hat{E} - \left( \ha + \frac{\sigma}{\hr} \right) x \; , \\
    \Sigma_\sigma &= \hr^2 \left( 1 - \frac{\sigma^2}{\hr^3} \right) \; , \\
    x &= \hat{J}_z - (\ha+\sigma)\hat{E} 
\end{align}
and
\begin{equation}
    J(\chi) = \sum_{k=0}^6 (1+e\cos\chi)^k \sum_{l=0}^k \frac{j^{(p)}_l j^{(e)}_{k-l}}{(1-e^2)^{k-l} p^l}
\end{equation}
with
\begin{align*}
    j^{(p)}_0 &= 1-\hat{E}^2 \; , \\
    j^{(p)}_1 &= -2 \; , \\
    j^{(p)}_2 &= \ha^2+2 \ha \hat{E} x+x^2 \; , \\
    j^{(p)}_3 &= -2((1-\hat{E}^2) \sigma^2 - \hat{E} \sigma x + x^2) \; , \\
    j^{(p)}_4 &= 4\sigma^2 \; , \\
    j^{(p)}_5 &= -2 \ha \sigma (\ha \sigma+x (\hat{E}\sigma +x)) \; , \\
    j^{(p)}_6 &= \sigma^2 ((1 - \hat{E}^2) \sigma^2 - 2 \hat{E} \sigma x - x^2) 
\end{align*}
and
\begin{align*}
    j^{(e)}_0 &= 1 \; , \\
    j^{(e)}_1 &= 2 \; , \\
    j^{(e)}_2 &= e^2+3 \; , \\
    j^{(e)}_3 &= 4 (e^2+1) \; , \\
    j^{(e)}_4 &= e^4+10 e^2+5 \; , \\
    j^{(e)}_5 &= 2 (e^2+3) (3 e^2+1) \; , \\
    j^{(e)}_6 &= e^6+21 e^4+35 e^2+7 \; .
\end{align*}

\section{Linearized evolution of the orbital parameters}
\label{app:linearized_inspiral}

In this Appendix we provide formulas for the evolution of the corrections $\delta p$ and $\delta e$ in Sec.~\ref{sec:EvolLin}. The evolution of the linear parts $\delta p$ and $\delta e$ is governed by Eqs.~\eqref{eq:ddeltapedt} where the functions $\delta\dot{p}$ and $\delta\dot{e}$ are
\begin{align}
    \delta \dot{p} &= \eval{\pdv{\dot{p}}{\sigma}}_{\sigma=0} + \pdv{\dot{p}^{\rm (g)}}{p} \delta p + \pdv{\dot{p}^{\rm (g)}}{e} \delta e \\
    \delta \dot{e} &= \eval{\pdv{\dot{e}}{\sigma}}_{\sigma=0} + \pdv{\dot{e}^{\rm (g)}}{p} \delta p + \pdv{\dot{e}^{\rm (g)}}{e} \delta e
\end{align}
After substitution from Eqs. \eqref{eq:dpdtdedt}, the $\sigma$ derivatives read
\begin{widetext}
\begin{align}
    \pdv{\dot{p}}{\sigma} &= \frac{ \displaystyle \pdv{\hat{J}_z}{e}{\sigma} \dot{\hat{E}} + \pdv{\hat{J}_z}{e} \delta \dot{\hat{E}} - \pdv{\hat{E}}{e}{\sigma} \dot{\hat{J}}_z - \pdv{\hat{E}}{e} \delta\dot{\hat{J}}_z}{\abs{J_{(\hat{E},\hat{J}_z)}}} - \frac{\displaystyle \pdv{\hat{J}_z}{e} \dot{\hat{E}} - \pdv{\hat{E}}{e} \dot{\hat{J}}_z}{\abs{J_{(\hat{E},\hat{J}_z)}}^2} \pdv{\abs{J_{(\hat{E},\hat{J}_z)}}}{\sigma} \\
    \pdv{\dot{e}}{\sigma} &= \frac{ \displaystyle-\pdv{\hat{J}_z}{p}{\sigma} \dot{\hat{E}} - \pdv{\hat{J}_z}{p} \delta \dot{\hat{E}} + \pdv{\hat{E}}{p}{\sigma} \dot{\hat{J}}_z + \pdv{\hat{E}}{p} \delta\dot{\hat{J}}_z}{\abs{J_{(\hat{E},\hat{J}_z)}}} - \frac{ \displaystyle -\pdv{\hat{J}_z}{p} \dot{\hat{E}} + \pdv{\hat{E}}{p} \dot{\hat{J}}_z}{\abs{J_{(\hat{E},\hat{J}_z)}}^2} \pdv{\abs{J_{(\hat{E},\hat{J}_z)}}}{\sigma} \\
    \pdv{\abs{J_{\textbf{}}}}{\sigma} &= \pdv{\hat{E}}{p}{\sigma} \pdv{\hat{J}_z}{e} + \pdv{\hat{E}}{p} \pdv{\hat{J}_z}{e}{\sigma} - \pdv{\hat{E}}{e}{\sigma} \pdv{\hat{J}_z}{p} - \pdv{\hat{E}}{e} \pdv{\hat{J}_z}{p}{\sigma}
\end{align}
\end{widetext}
where $\delta \dot{E}$ and $\delta \dot{J}_z$ are given by the linear parts of the fluxes
\begin{align}
    \delta\dot{\hat{E}} &= - q \eval{\delta \mathcal{F}^{E}}_{p,e} \\
    \delta\dot{\hat{J}}_z &= - q \eval{\delta \mathcal{F}^{J_z}}_{p,e}
\end{align}

The derivatives of $\dot{p}$ and $\dot{e}$ with respect to $p$ and $e$ are calculated similarly, while the derivatives of the constants of motion with respect to $p$, $e$ and $\sigma$ can be calculated from Eqs.~\eqref{eq:energy2} and \eqref{eq:angmom2}. The exact formulas of the latter are not presented here, because even if they are straightforward to calculate, they have long complex forms.

\section{Linearized partial amplitudes}
\label{app:linearized_amplitudes}

Here we give more details about the calculation of the linearized in spin partial amplitudes $\delta C^\pm_{lmn}$ (Eq.~\eqref{eq:deltaCpm_lmn}). The linear part of $I^\pm_{lmn}$ from Eq. \eqref{eq:Ipm} reads
\begin{widetext}
\begin{equation}
    \dv{I^\pm_{lmn}}{\sigma} = \frac{1}{W} \qty( \delta A_0 - \qty( \delta A_1 + \delta B_0 - A_0 \delta r )\dv{r} + \qty( \delta A_2 + \delta B_2 - A_1 \delta r )\dv[2]{r} - \qty( \delta B_3 - A_2 \delta r ) \dv[3]{r} ) R^\mp_{lm\omega} \, ,
\end{equation}
\end{widetext}
where the coefficients $\delta A_i$ and $\delta B_i$ are calculated by the linearization in spin of the expressions in Eqs.~(B1-B3) and (B9-B11) of the Appendix B of \citep{Skoupy:2021b}. Particularly, the linear part of $A^0_{abi}$ is calculated as
\begin{equation}
    \delta A_{abi}^{0} = \left( \delta C_{ab}^{0} - \delta C_{ab}^{\sigma} \right) f_{ab}^{(i)} + C_{ab}^{0 {\rm (g)}} \dv{f{}_{ab}^{(i)}}{r} \delta r \, ,
\end{equation}
and the calculation of $\delta A^{t\phi}_{abi}$, $\delta A^r_{abi}$ and $\delta B_i$ is trivial because these functions are proportional to $\sigma$.

\begin{figure}
    \centering
    \includegraphics[width=0.45\textwidth]{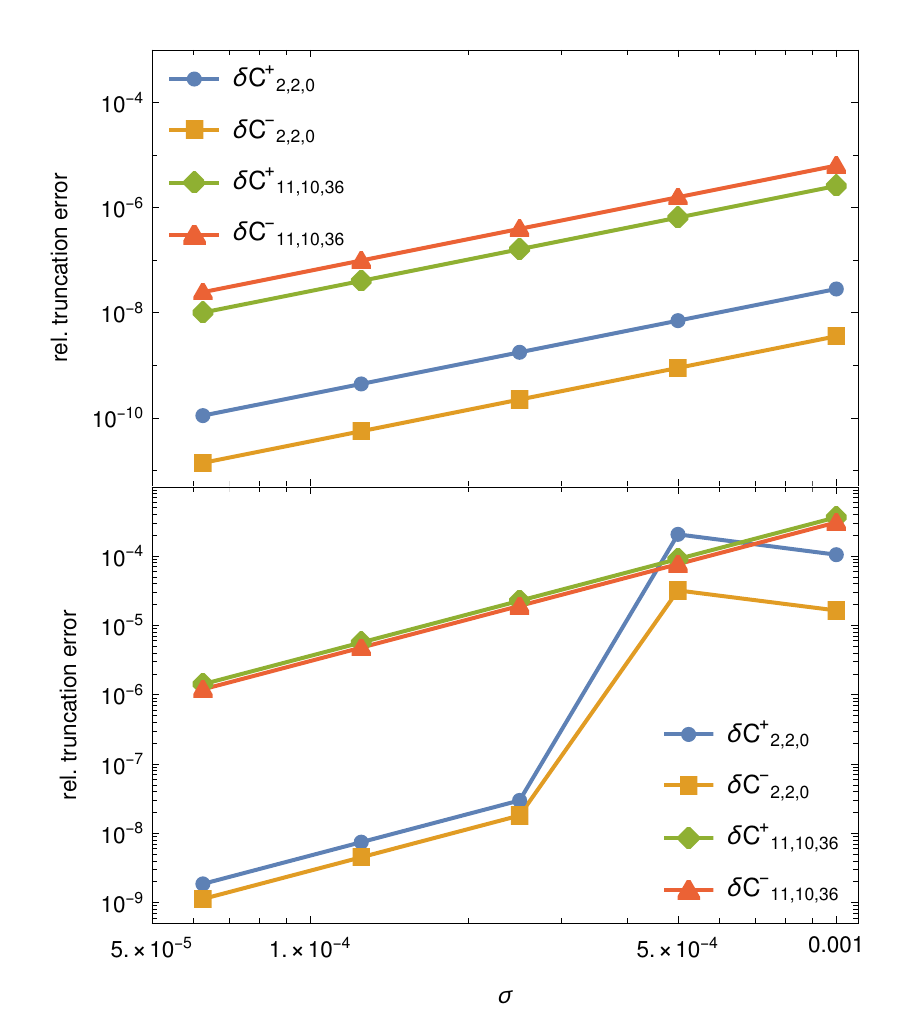}
    \caption{The relative truncation errors~\eqref{eq:truncation} for $\ha=0.9$, $p^{\rm (g)}=12$, $e^{\rm (g)}=0.6$ (top) and $\ha=0.9$, $p^{\rm (g)}=4$, $e^{\rm (g)}=0.4$ (bottom). These errors tend to zero for sufficiently small $\sigma$ and the calculation of $\delta C^{\pm}_{lmn}$ is therefore correct.}
    \label{fig:truncation}
\end{figure}

The linear parts of the partial amplitudes $\delta C^\pm_{lmn}$ are calculated simultaneously with the geodesic amplitudes $C^{\pm {\rm (g)}}_{lmn}$. We have tested the results against non-linearized partial amplitudes $C^\pm_{lmn}$ by comparing them with numerical $\sigma$ derivatives of $C^\pm_{lmn}$ with respect to a reference geodesic with the same frequencies. To find the orbital parameters of a trajectory of a spinning particle with the same frequencies as those of a geodesic with $p^{\rm (g)}$ and $e^{\rm (g)}$, we numerically calculated $p^\pm$ and $e^\pm$ satisfying
\begin{equation}
    \Omega_i(p^\pm,e^\pm,\pm \sigma) = \Omega^{\rm (g)}_i(p^{\rm (g)},e^{\rm (g)})
\end{equation}
Then we numerically calculated the derivative
\begin{equation}
    \delta C^{\pm {\rm Num}}_{lmn} = \frac{C^\pm_{lmn}(p^+,e^+,\sigma)-C^\pm_{lmn}(p^-,e^-,-\sigma)}{2\sigma}
\end{equation}
and the relative difference
\begin{equation}\label{eq:truncation}
    \abs{1-\frac{\delta C^{\pm {\rm Num}}_{lmn}}{\delta C^\pm_{lmn}}}
\end{equation}
If the calculation of $\delta C^\pm_{lmn}$ from Eq.~\eqref{eq:deltaCpm_lmn} is correct, then the relative difference equals to the relative truncation error of second order finite difference formula and behaves as $\order{\sigma^2}$.

We have calculated the relative difference for two orbits, namely with $p^{\rm (g)}=12$, $e^{\rm (g)}=0.6$ and $p^{\rm (g)}=4$, $e^{\rm (g)}=0.4$ for $\ha=0.9$ and for two modes with $l=2$, $m=2$, $n=0$ and $l=11$, $m=10$, $n=36$. The results are plotted in Fig.~\ref{fig:truncation}. We can see that for sufficiently small $\sigma$ the relative error tends to zero and, therefore, the linear parts $\delta C^\pm_{lmn}$ are correct.

\section{Accuracy of the interpolation}
\label{app:accuracyInt}

In this Appendix we discus the interpolation error originated when interpolating the fluxes and other quantities in the $p-e$ plane in Section \ref{sec:interpolation}.

We use global interpolation on the Chebyshev nodes. The advantage of this method is that the convergence is exponential and the interpolation error is bounded and uniform. The disadvantage is that the convergence is slow, when the function is not analytical, and the errors in the evaluation at individual points spread across the whole domain.

\begin{figure}
    \centering
    \includegraphics[width=0.45\textwidth]{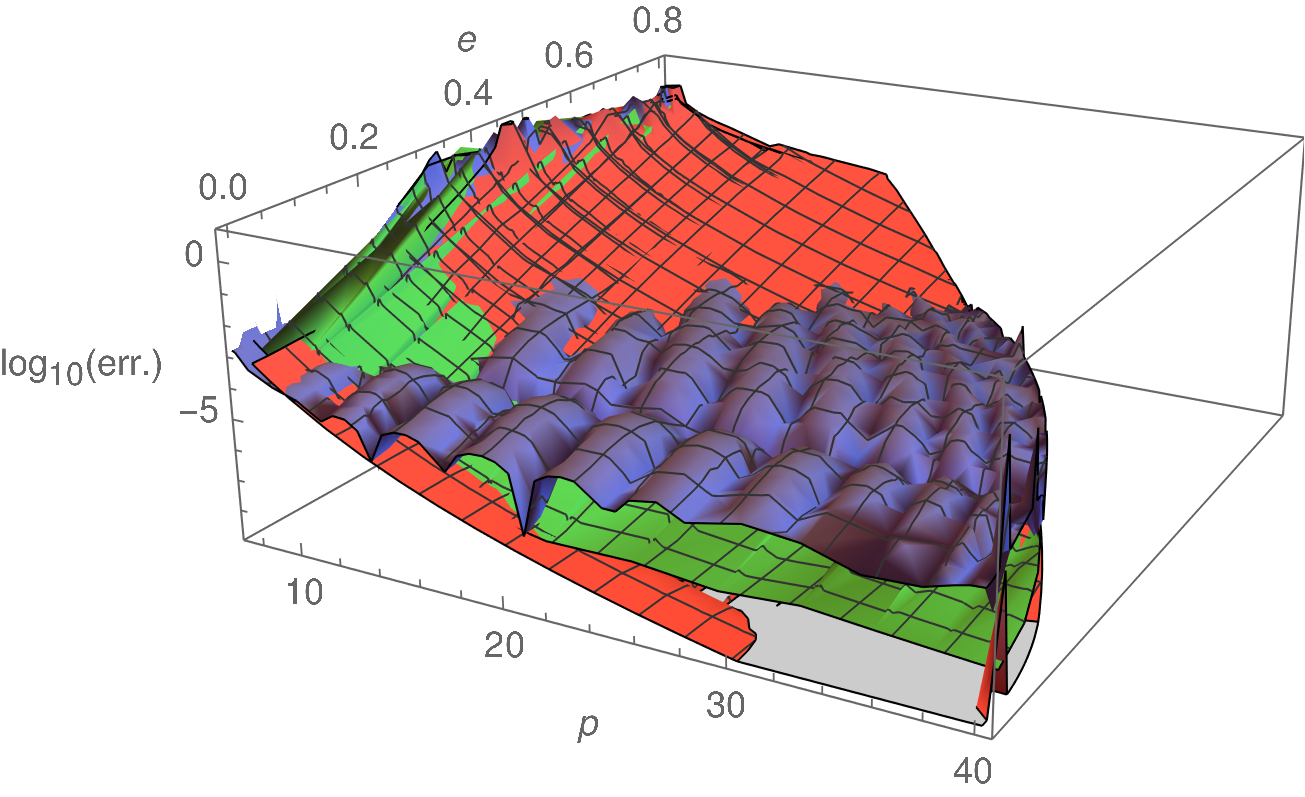}
    \caption{The relative error of the interpolated energy flux to infinity compared to a 9PN series (purple), the relative error at individual grid points (green) and the relative error of the 9PN series deducted from the last term (red). We can see that the relative interpolation error is around $10^{-4}$. In the area near the separatrix or with high eccentricity the 9PN series loses accuracy.}
    \label{fig:intError}
\end{figure}

The interpolation error of the Chebyshev interpolation can be easily estimated. Namely, when a function $f$ is expanded into the Chebyshev polynomials as
\begin{equation}
    f(x,y) = \sum_{i=1}^{i_{\rm max}} \sum_{j=1}^{j_{\rm max}} c_{ij} T_i(x) T_j(y) \; ,
\end{equation}
where $T_i(x)$ are Chebyshev polynomials and $c_{ij}$ are the coefficients, then the error can be estimated as
\begin{equation}
    \max_{ i = i_{\rm max} \vee j = j_{\rm max} } \abs{c_{ij}} \, .
\end{equation}
Using this approach we have found that the relative error of the interpolated geodesic fluxes $\mathcal{F}^{\rm (g)}$ is around $10^{-4}$, the relative error of $\dot{p}^{\rm (g)}$ and $\dot{e}^{\rm (g)}$ is around $10^{-5}$ and the relative error of the derivatives of $\dot{p}$ and $\dot{e}$ is between $10^{-3}$ and $10^{-2}$. Since the functions $\delta\dot{p}$ and $\delta\dot{e}$ are calculated from these derivatives, their precision is also between $10^{-3}$ and $10^{-2}$.

To verify the geodesic energy flux to infinity for the Schwarzschild black hole we compared the data with 9PN series \citep{BHPToolkit}. Fig.~\ref{fig:intError} shows both relative difference between the PN series and the interpolated function and the value of the flux at individual points. It also shows the error of the PN series estimated by its last term. We can see that the interpolation error is dominant for higher $p$ and lower $e$ and its value is around $10^{-4}$. The error of the fluxes at individual grid points is between $10^{-8}$ and $10^{-7}$, but the error of the PN series grows with decreasing $p$ and increasing $e$ and, therefore, the fluxes near the separatrix cannot be verified using the PN series.

\section{Accuracy of the phase shifts}
\label{app:accuracyPhases}

In this Appendix we compare the linearized phase shifts $\delta \Phi_i(t)$ obtained in Sec.~\ref{sec:EvolLin} with the phases computed using non-linearized formula \eqref{eq:Phase}. The purpose of this section is to test the validity and accuracy of the calculation. 

First we have computed the non-linearized fluxes on a grid in the $p-e$ plane for $\sigma=10^{-3}$ and $\ha=0$. The grid is similar to the grid for $\ha=0$ in Fig.~\ref{fig:grids}, but the separatrix is located at different position fulfilling $p_{\rm s}(\sigma) = 6+2e+\order{\sigma}$, i.e. around $10^{-3}$ away from the geodetic separatrix. The calculation of the non-linearized fluxes was equivalent to the calculation of linearized fluxes in Sec.~\ref{sec:implementation}.

\begin{figure}[b!]
    \centering
    \includegraphics[width=0.45\textwidth]{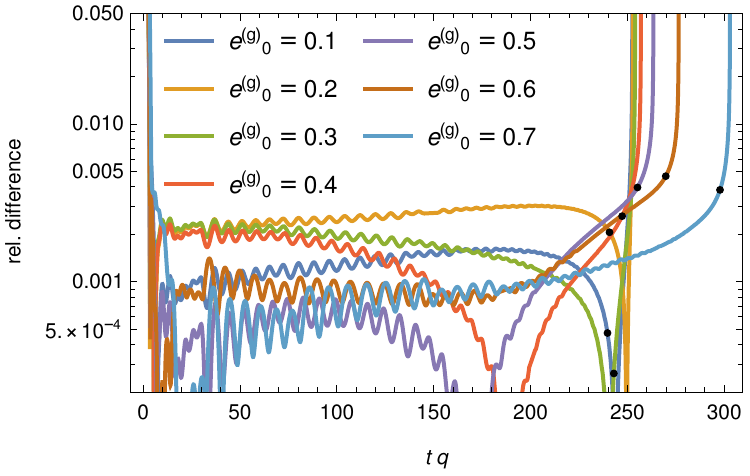}
    \includegraphics[width=0.45\textwidth]{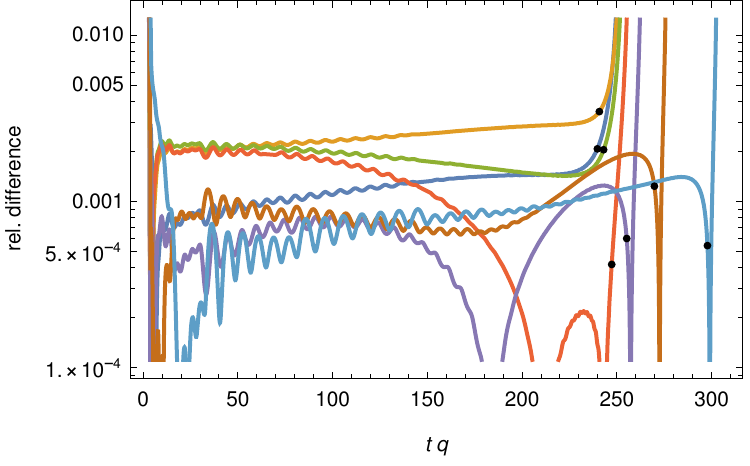}
    \caption{The relative difference between the non-linearized phase shift $\Delta\Phi_i$ and the linearized phase shift $\delta \Phi_i$ for $i=\phi$ (top) and $i=r$ (bottom). It can be seen that the relative difference is below $5\times 10^{-3}$ for the majority of the inspiral. At the end of the inspiral the relative difference grows rapidly because the linearization in $\sigma$ breaks. The black dots show points where $\delta\Phi_r$ changes from increasing to decreasing function of $t$ and it has maximal value. The error at these points is below $5\times 10^{-3}$.}
    \label{fig:ddPhi_i}
\end{figure}

We have computed the evolution of the orbital parameters $p(t)$ and $e(t)$ using Eqs.~\eqref{eq:dpdtdedt} and from $p(t)$ and $e(t)$ we calculated the phases \eqref{eq:Phase}. The initial orbital parameters $p_0$ and $e_0$ were chosen to match the initial frequencies of a geodesic with initial parameters $p_0^{\rm (g)}$ and $e_0^{\rm (g)}$. Similarly we have calculated the phase for $\sigma=0$. We have compared the phase shift
\begin{equation}
    \Delta \Phi_i = \frac{\Phi_i(\sigma) - \Phi_i(\sigma=0)}{\sigma}
\end{equation}
with the linear part of the phase $\delta \Phi_i$ as
\begin{equation}
    \abs{ 1-\frac{\delta\Phi_i}{\Delta\Phi_i} }\, .
\end{equation}
This relative difference is plotted in Fig.~\ref{fig:ddPhi_i} for initial semi-latus rectum $p^{\rm (g)}_0=12$ and different initial eccentricities $e^{\rm (g)}_0$. We can see that the relative difference is below $5\times 10^{-3}$ for the majority of the inspiral. Before the particle reaches the separatrix, the relative difference diverges because the liearization in $\sigma$ breaks here. This is caused by the fact that the linear parts $\delta p(t)$ and $\delta e(t)$ diverge here and the functions as $\dot{p}(p^{\rm (g)}(t) + \sigma \delta p(t), e^{\rm (g)}(t) + \sigma \delta e(t), \sigma)$ cannot be linearized.

Since the quantity $\Delta \Phi_i$ is non-linearized, it contains $\order{\sigma}$ contribution to the phase which should be around $10^{-3}$. However, since the accuracy of the calculations is around $10^{-3}$, the relative difference shows this numerical error.

\bibliography{paper}

\end{document}